\documentclass[fleqn,usenatbib]{mnras}

\usepackage{newtxtext,newtxmath}

\usepackage[T1]{fontenc}

\DeclareRobustCommand{\VAN}[3]{#2}
\let\VANthebibliography\thebibliography
\def\thebibliography{\DeclareRobustCommand{\VAN}[3]{##3}\VANthebibliography}

\usepackage{graphicx}	
\usepackage{amsmath}	
\usepackage{mhchem}
\usepackage{xcolor}
\DeclareMathOperator{\Var}{\operatorname{Var}}
\DeclareMathOperator{\E}{\operatorname{E}}

\title[Clouds on GJ~1214b]{The Impact of Phase Equilibrium Cloud Models on GCM Simulations of GJ~1214b}

\author[D. A. Christie et al.]{
D. A. Christie,$^{1}$\thanks{E-mail: d.christie@exeter.ac.uk}
N. J. Mayne,$^{1}$
R. M. Gillard,$^{1}$
J. Manners,$^{2}$
E. Hébrard,$^{1}$
S. Lines,$^{2}$
and K. Kohary$^{1}$
\\
$^{1}$Physics and Astronomy, College of Engineering, Mathematics and Physical Sciences, University of Exeter, Exeter EX4 4QL, UK \\
$^{2}$Met Office, FitzRoy Road, Exeter EX1 3PB, UK
}

\date{Accepted XXX. Received YYY; in original form ZZZ}

\pubyear{2022}

\begin{document}
\label{firstpage}
\pagerange{\pageref{firstpage}--\pageref{lastpage}}
\maketitle

\begin{abstract}
We investigate the impact of clouds on the atmosphere of GJ~1214b using the radiatively-coupled, phase-equilibrium cloud model {\sc EddySed} coupled to the {\sc Unified Model} general circulation model. We find that, consistent with previous investigations, high metallicity ($100\times$ solar) and clouds with large vertical extents (a sedimentation factor of $f_\mathrm{sed} = 0.1$) are required to best match the observations, although metallicities even higher than those investigated here may be required to improve agreement further.  We additionally find that in our case which best matches the observations ($f_\mathrm{sed}=0.1$), the velocity structures change relative to the clear sky case with the formation of a superrotating jet being suppressed, although further investigation is required to understand the cause of the suppression.  The increase in cloud extent with $f_\mathrm{sed}$ results in a cooler planet due to a higher albedo, causing the atmosphere to contract. This also results in a reduced day-night contrast seen in the phase curves, although the introduction of cloud still results in a reduction of the phase offset.  We additionally investigate the impact the the {\sc Unified Model}'s pseudo-spherical irradiation scheme on the calculation of heating rates, finding that the introduction of nightside shortwave heating results in slower mid-latitude jets compared to the plane parallel irradiation scheme used in previous works.  We also consider the impact of a gamma distribution, as opposed to a log-normal distribution, for the distribution of cloud particle radii and find the impact to be relatively minor. 
\end{abstract}




\section{Introduction}
GJ~1214b, the archetypal warm Neptune, was one of the first mini-Neptunes/super-Earths to be discovered \citep{charbonneau_2009} and is notable for its featureless transmission spectrum \citep{kreidberg_2014,gillon_2014}. It has been the focus of general circulation model (GCM) simulations for a number of years \citep[e.g.,][]{menou_2012,kataria_2014,charnay_2015a,charnay_2015b, drummond_2018}, and has served as a test case for the limits of applicability of the primitive equations in mini-Neptunes \citep{mayne_2019} as well as tests of model convergence \citep{wang_2020}. Despite the inference of clouds in its atmosphere \citep{kreidberg_2014,gillon_2014}, GCM simulations of GJ~1214b have mostly ignored clouds, leaving their investigation to one dimensional models.  The exception to this is \citet{charnay_2015b} which modelled the atmosphere of GJ~1214b including clouds consisting of radiatively-active cloud droplets of fixed radii.  They found that particles of $0.5\,\mathrm{\mu m}$ in size could be lofted high in the atmosphere and their impact on the transmission spectrum improved agreement with observations.  

While not capturing the complete dynamics, one dimensional models allow for improved modelling of the chemistry and cloud physics. \citet{morley_2013} modelled GJ~1214b with a photochemical kinetics model with a parametrised vertical mixing set through the eddy mixing rate, $K_{zz}$, including both \ce{KCl} and \ce{ZnS} clouds as well as hazes, and found that enhanced metallicity could explain the observations if the sedimentation efficiency was sufficiently low to result in large cloud scale heights.  They also found that photochemical haze, with $1\%$ to $5\%$ of haze precursors (e.g., \ce{C_2H_2}, \ce{C_2H_4}, etc.) forming haze particles, could explain the observations.  The latter mechanism does not require atmospheric mixing to carry the particles high into the atmosphere as hazes are expected to form at the low pressures needed to explain observations, and the small particle sizes and associated sedimentation timescales allow hazes to persist in the upper atmosphere.  More sophisticated modeling of the cloud formation process has been done using the {\sc CARMA} cloud microphysics code by \citet{gao_2018} who modelled \ce{KCl} and \ce{ZnS} clouds on GJ~1214b.  They found that, in the absence of haze, metallicities of $1000\times $ solar and mixing rates of $K_{zz}=10^{6}\,\mathrm{m^2s^{-1}}$ are required to adequately explain the flat transmission spectrum.

The modelling of clouds in three dimensions often takes the form of post-processing GCM models due to the simplicity and relative computational ease \citep[e.g.,][]{robbins-blanch_2022}, allowing for the generation of synthetic observables; however, without the self-consistent coupling of the cloud layer to the atmosphere, the full impact of clouds may be missed. GCM simulations of radiatively-coupled clouds on gas giants have primarily focused on hot Jupiters and have been done adopting a range of complexities, from modelling the cloud formation microphysics \citep[e.g.,][]{lee_2016, lines_2018a} to the use of parametrised cloud models \citep[e.g.,][]{lines_2019,christie_2021} or fixed particle sizes \citep[e.g.,][]{charnay_2015b,parmentier_2020,roman_2021}. Although the increased complexity comes at computational expense compared to post-processing, these models find that radiatively-coupled clouds can have an impact on the thermal structure of the atmosphere, motivating their necessity.

In this paper, we study clouds in the atmosphere of GJ~1214b using the phase-equilibrium {\sc EddySed} cloud model radiatively coupled to the Met Office's {\sc Unified Model} (UM) GCM.  The use of a one dimensional could model coupled to a GCM provides a complementary view of cloud formation to \citet{charnay_2015b} in that while it does not couple the advection and sedimentation directly to the atmosphere via tracers, it allows for a size distribution to be modelled with the particle sizes varying throughout the atmosphere.  We use this setup to investigate the impact of clouds on both the atmospheric dynamics as well as the synthetic observations. Overall, we find that supersolar metallicities and large cloud scale heights are required to significantly impact the dynamics and the observables, consistent with previous studies, and that this increased cloud also results in a cooling and contraction of the atmosphere due to the increased albedo.

The structure of the paper is as follows: In Section \ref{Sec:Model}, we outline the components of the simulations. In Section \ref{Sec:Results}, we present the result from the simulations with conclusions and a summary found in Section \ref{Sec:Conclusions}. We also include a number of appendices. In Appendix \ref{Apdx:SphIrr}, we demonstrate the impact of switching from a plane-parallel zenith-angle approximation in attenuating incoming stellar radiation to a more physically motivated pseudo-spherical geometry method. In Appendix \ref{Apdx:Gamma}, we explore the impact of particle sizes being distributed in a gamma distribution instead of a log-normal distribution.  In Appendix \ref{Apdx:Cond}, we investigate the impact of \ce{KOH} on \ce{KCl} cloud formation.  In Appendix \ref{Apdx:Conv}, we discuss the convergence properties of the simulations presented in the paper. In Appendix \ref{Apdx:Plots}, we present a number of additional plots.

\section{The Model}
\label{Sec:Model}
To understand the impact of clouds on the atmosphere of GJ~1214b, we simulate the atmosphere using the {\sc UM} coupled to the {\sc EddySed} phase-equilibrium cloud model.  We introduce these models and present the specifics of the simulations below.

\subsection{The {\sc Unified Model} GCM}

The {\sc UM} solves the full, deep-atmosphere, non-hydrostatic Navier–Stokes equations \citep[][]{mayne_2014a,wood_2014} and has been adapted to model hot Jupiters \citep{mayne_2014a,mayne_2017} and mini-Neptunes \citep{drummond_2018,mayne_2019}.  Radiative transfer is done using the {\sc Socrates} radiative transfer code based on \citet{edwards_1996} which has been adapted and benchmarked in \citet{amundsen_2014a,amundsen_2017}. 
We assume an atmosphere dominated by \ce{H_2} and \ce{He} with the gas-phase opacity sources being \ce{H_2O}, \ce{CO}, \ce{CO_2}, \ce{CH_4}, \ce{NH_3}, \ce{Li}, \ce{Na}, \ce{K}, \ce{Rb}, \ce{Cs} and \ce{H_2}-\ce{H_2} and \ce{H_2}-\ce{He} collision-induced absorption (CIA). Opacities are computed using the correlated-k method and {\sc ExoMol} line lists \citep{tennyson_2012,tennyson_2016}.  Unless discussed below, the setup is the same as our previous simulations using a coupled version of {\sc EddySed} \citep{christie_2021} except with the planetary parameters appropriate for GJ~1214b. The common parameters used across all simulations are found in Table \ref{Tbl:Common}.

Previous studies which coupled {\sc EddySed} to the {\sc UM} \citep{lines_2018a,christie_2021} did not include $\mathrm{CO_2}$ as a source for gas-phase opacity as it has a relatively low abundance in atmospheres with solar metallicity; however, as mini-Neptune atmospheres may have enhanced metallicities, meaning $\mathrm{CO_2}$ abundances may become significant, we follow \citet{charnay_2015a} in taking the abundances of \ce{CO}, \ce{CO_2}, \ce{H_2O}, and \ce{CH_4} to be in chemical equilibrium through the reactions
\begin{align}
    &\mathrm{CO} + 3\mathrm{H_2} \rightleftharpoons \mathrm{H_2O} + \mathrm{CH_4} \nonumber \,\,,\\
    &\mathrm{CO_2} + \mathrm{H_2} \rightleftharpoons \mathrm{H_2O} + \mathrm{CO}\,\, ,
\end{align}
\noindent with the solution found using a modification of the analytic chemistry method presented in the appendix of \citet{burrows_1999}.    

Due to the uncertainty in atmospheric composition of GJ~1214b, we consider two different metallicities, solar and $100\times$ solar, with a focus on the $100\times$ solar case.  As differing atmospheric metallicities can result in very different atmospheric scale heights, thus requiring different computational domain heights, we situate each computational domain such that the midpoint of the vertical axis is located at the observed planetary radius ($1.7\times 10^7\,\mathrm{m}$). While this does result in differing gravitational accelerations $g$ at the inner boundary, as the {\sc UM} accounts for spatial variation in $g$, the gravitational accelerations in regions where the two domains overlap agree. \citet{gao_2018} found that a metallicity of $1000\times$ solar best fit the observations. For such a high metallicity, the assumption that the atmosphere is dominated by hydrogen and helium breaks down, and properly accounting for these atmospheric compositions requires modifications to the {\sc Socrates} configuration. These higher metallicites may be investigated in a future work, but are beyond the scope of the investigation presented here.

To compute the gas constant $R$, heat capacity $c_p$, and mean molecular weight, we compute equilibrium chemistry profiles for GJ~1214b for solar and $100\times$ solar metallicities using the 1D radiative-convective code {\sc Atmo} \citep{tremblin_2015,drummond_2016} and average the properties over the model atmosphere (see Table \ref{Tbl:GasProp}).  These 1D profiles also serve as the initial conditions for our simulations.

We also make use of the {\sc UM}'s new pseudo-spherical irradiation scheme \citep{jackson_2020}.  In our previous works \citep[e.g.,][]{christie_2021}, a plane-parallel scheme was used for attenuating the incoming shortwave radiation.  When using the old scheme, the incoming shortwave radiation goes to zero at the terminator, precluding any possibility of nightside heating. The new pseudo-spherical scheme remedies this by computing the attenuation using spherical shells.  We find that in allowing for nightside heating  there is a reduction of wind speed near the poles which may be physically relevant in scenarios where there is a mid-latitude or polar jet.  We present the results from our tests of the new scheme in Appendix \ref{Apdx:SphIrr}. 

The radiative timestep for most simulations is $150\,\mathrm{s}$; however, in the case $100\times$ solar metallicity simulations with $f_\mathrm{sed}=0.5$ and $0.1$ the increased opacity in the upper atmosphere necessitates a shorter radiative timestep of $60\,\mathrm{s}$. 
\begin{table}
\caption{Common Parameters}
\label{Tbl:Common}
\begin{tabular}{lc}
\hline
  & Value \\
\hline
{\em Grid and Timestepping} \\
Longitude Cells & 144 \\
Latitude Cells & 90 \\
Hydrodynamic Timestep & 30 s \\
\\
{\em Radiative Transfer} \\
Wavelength Bins & 32 \\
Wavelength Minimum & 0.2 $\mathrm{\mu m}$ \\
Wavelength Maximum & 322 $\mathrm{\mu m}$ \\
\\
{\em Damping and Diffusion} \\
Damping Profile & Polar \\
Damping Coefficient & 0.2 \\
Damping Depth ($\eta_s$) & 0.7 \\
Diffusion Coefficient & 0.158 \\
\\
{\em Planet}\\
Intrinsic Temperature & 100 K \\
Initial Inner Boundary Pressure & 200 bar \\
Semi-major axis $a$ & $1.23\times 10^{-2}$ AU \\
Stellar Constant (at 1 AU) & 3.9996 $\mathrm{W\, m^{-2}}$ \\
\hline
\end{tabular}
\end{table}

\begin{table}
\caption{Metallicity Dependent Parameters}
\label{Tbl:GasProp}
\begin{tabular}{lcc}
\hline
  & Solar & $100\times$ Solar \\
\hline
{\em Gas Parameters} \\
Specific Gas Constant R (J/kg/K) & $3.513\times 10^3$ & $1.565\times 10^3$ \\
Specific Heat Capacity $c_p$ (J/kg/K) & $1.200\times 10^4$    & $5.632\times 10^3$    \\
Mean Molecular Weight (g/mol) & $2.36$ & $5.31$ \\ 
\\
{\em Simulation Parameters} \\
Inner boundary (m) & $1.46\times 10^7$ & $1.58\times 10^7$\\
Domain height (m)  & $4.8\times 10^6$ & $2.4\times 10^6$\\
$g$ at inner boundary ($\mathrm{m/s^2}$) & $12.2$ & $10.4$ \\
$K_{zz,0}$ ($\mathrm{m^2}/s$) & $7\times 10^2$ & $3\times 10^3$ \\
Radiation Timestep (s) & $150$ & $60$/$150$ \\
\hline
\end{tabular}
\end{table}

\subsection{The {\sc EddySed} Cloud Model}
To include clouds in our simulations, we couple to the {\sc UM} the one-dimensional, phase-equilibrium, parametrised cloud model {\sc EddySed} \citep{ackerman_2001}, treating each vertical column within the GCM independently. The benefit of this approach is that it allows for a level of sophistication beyond the assumption of fixed cloud decks and particle sizes, while still being fast enough to allow long simulation times.  The underlying assumption is that clouds exist within an equilibrium where vertical mixing is balanced by sedimentation, 
\begin{equation}
\frac{\partial q_\mathrm{t}}{\partial z} = -\frac{\left<v_\mathrm{sed} \right>}{K_{zz}}q_\mathrm{c} ,
\label{Eqn:Balance}
\end{equation}
\noindent where $q_\mathrm{c}$ is the condensate mass mixing ratio, $q_\mathrm{v}$ is the vapour mass mixing ratio, and $q_\mathrm{t} = q_\mathrm{c} + q_\mathrm{v}$.  {\sc EddySed} further assumes that the mass-averaged sedimentation velocity $\left<v_\mathrm{sed}\right>$ is proportional to a characteristic mixing velocity $w_\star = K_{zz}/L_\mathrm{mix}$, where $L_\mathrm{mix}$ is the characteristic mixing scale, which we take to be equal to the pressure scale height $H$ and the constant of proportionality is $f_\mathrm{sed}$.\footnote{As discussed in \citet{ackerman_2001} where the {\sc EddySed} model is applied to brown dwarfs, in convective atmospheres the characteristic scale $L_\mathrm{mix}$ is taken to be the convective mixing scale.  As we are applying the model to radiation-dominated regions of the atmosphere, we opt for the simpler assumption of $L_\mathrm{mix}=H$.}   Combined with Equation \ref{Eqn:Balance}, this yields the governing equation
\begin{equation}
\frac{\partial q_\mathrm{t}}{\partial z} =  -f_\mathrm{sed}\frac{w_\star}{K_{zz}} q_\mathrm{c} = -\frac{f_\mathrm{sed}}{L_\mathrm{mix}}q_\mathrm{c}\,\, 
\end{equation}
\noindent where $z$ is the vertical coordinate. The cloud distribution is thus determined by a cloud scale height $L_\mathrm{cloud}=f_\mathrm{sed}^{-1}L_\mathrm{mix}$ and is independent of $K_{zz}$, except through the impact of $K_{zz}$ on the particle sizes and thus the heating rates. It is assumed that any vapour in excess of the local vapour saturation mass mixing ratio $q_\mathrm{vs}$ immediately becomes condensate,
\begin{equation}
    q_\mathrm{c} = \max(0,q_\mathrm{t} - q_\mathrm{vs})\,\, .
\end{equation}

The sedimentation factor $f_\mathrm{sed}$ is taken to be constant, as in \citet{lines_2018a,christie_2021}; however, we note that \citet{rooney_2021} have investigated variations of the {\sc EddySed} model with an altitude dependent $f_\mathrm{sed}$.  Unfortunately, as the model adopted in \citet{rooney_2021} requires parameters that are poorly constrained for GJ~1214b, we opt to retain $f_\mathrm{sed}$ as a fixed parameter subject to a parameter study.  

Particle sizes are assumed to be log-normal with the peak of the distribution $r_\mathrm{g}$ given by
\begin{equation}
    r_\mathrm{g} = r_w f_\mathrm{sed}^{1/\alpha}\exp\left(-\frac{\alpha + 6}{2}\ln^2\sigma\right),
\end{equation}
\noindent where $\sigma$ parametrises the distribution width, $r_w$ is the particle radius at which the sedimentation speed is equal to $w_\star$ ($v_\mathrm{sed}(r_w) = w_\star=K_{zz}/H$), and $\alpha$ is the scaling of $v_\mathrm{sed}$ with particle radius, $v_\mathrm{sed}~\propto~r^\alpha$, estimated at $r=r_w$.  While $r_\mathrm{g}$ corresponds to the peak of the distribution, the area-weighted effective radius $r_\mathrm{eff}$, more relevant to radiative properties, is given by
\begin{equation}
    r_\mathrm{eff} = r_w f_\mathrm{sed}^{1/\alpha}\exp\left(-\frac{\alpha+1}{2}\ln^2\sigma\right).
    \label{Eqn:reff}
\end{equation}
The assumption of log-normality of the radius distribution results in the exponential dependence of $r_\mathrm{g}$ and $r_\mathrm{eff}$ on model parameters. As a point of comparison, we derive the {\sc EddySed} particle sizes for a gamma distribution and investigate the impact of such a choice in Appendix \ref{Apdx:Gamma}.  

The particle sedimentation speed $v_\mathrm{sed}$, used in the computation of $r_w$, is computed in the terminal velocity limit using
\begin{equation}
    v_\mathrm{sed}(r) = \frac{2}{9}\frac{\beta g r^2\Delta \rho}{\eta}\,\, ,
\end{equation}
\noindent where $\beta = 1 + K_\mathrm{n}\left(1.256+0.4\exp(-1.1/K_\mathrm{n}) \right)$ is the Cunningham slip factor, $K_\mathrm{n} = \lambda/r$ is the Knudsen number, $\lambda$ is the mean free path, $g$ is the gravitational acceleration, $\Delta \rho = \rho_\mathrm{c} - \rho_\mathrm{a}$ is the relative density of the cloud particles, and 
\begin{equation}
    \eta = \frac{5}{16}\frac{\sqrt{\pi m k_\mathrm{B} T}\left(k_\mathrm{B}T/\epsilon\right)^{0.16}}{1.22\pi d^2}\,\, 
\end{equation}
\noindent is the dynamic viscosity within the atmosphere \citep{rosner_2000}.  The molecular diameter of \ce{H_2} is $d = 2.827\times 10^{-8}\,\mathrm{cm}$, $\epsilon=59.7 k_\mathrm{B}\,\mathrm{K}$ is the depth of the Lennard-Jones potential well, $k_\mathrm{B}$ is the Boltzmann constant, and $m$ is the mean mass of a particle in the gas phase within the atmosphere.

As in \citet{christie_2021}, we assume the mixing rate $K_{zz}$ can adequately be described by the global average mixing.  For GJ~1214b, we adopt mixing rates from \citet{charnay_2015a} who performed a GCM tracer study and fit the resulting distribution to a one-dimensional analytic model to estimate $K_{zz}$,
\begin{equation}
K_{zz} = \frac{K_{zz,0}}{P^{0.4}_\mathrm{bar}},
\end{equation}
\noindent where $P$ is the pressure and the specific value of $K_{zz,0}$ is metallicity dependent (see Table \ref{Tbl:GasProp}).   The introduction of clouds may alter the structure of the atmosphere and thus the mixing; however, in our previous study of HD~209458b, we did not find the mixing to be significantly impacted \citep{christie_2021}.

The critical vapour saturation mixing ratios $q_\mathrm{vs}$ are adopted from \citet{morley_2012} and \citet{visscher_2006}. We only consider $\mathrm{KCl}$ and $\mathrm{ZnS}$ as most other condensable species are expected to have condensed near the base of or below our computational boundary.  \ce{KCl} clouds are expected to condense directly from gas phase \ce{KCl}, which at solar metallicities, is the dominant \ce{K}-bearing species in the gas phase.  The critical saturation vapour pressure is given by
\begin{equation}
    P_\mathrm{vs,KCl} = 10^{7.6106 - 11382\, \mathrm{K}/T}\,\,\mathrm{bar},
\end{equation}
\noindent following \citet{morley_2012}. At metallicities of $100\times$ solar, however, \ce{KOH} becomes more abundant in the gas phase than \ce{KCl} in the limit of chemical equilibrium. To ascertain the of \ce{KOH} impact on cloud formation, we examine \ce{KCl} condensation in the presence of \ce{KOH} using {\sc Atmo} in Appendix \ref{Apdx:Cond}. We find that the inclusion of \ce{KOH} does shift the cloud deck slightly; however, once cloud begins to form, both gas-phase \ce{KCl} and \ce{KOH} are quickly depleted and the cloud mixing ratio is insensitive to the inclusion of \ce{KOH}.  

Unlike \ce{KCl}, \ce{ZnS} is not expected to exist in the gas phase and as a result forms via the chemical reaction
\begin{equation}
\ce{Zn + H_2S -> H_2 + ZnS(s)}
\end{equation}
\noindent directly into the solid state. Due to high surface energies, \ce{ZnS} clouds likely require nucleation sites as precursors to cloud formation \citep{gao_2018}.  Within our models we assume that such sites exist and ultimately do not impact the final distribution of cloud particles.  We adopt a condensation curve from \citet{visscher_2006,morley_2012}, including a correction for the atmospheric metallicity, namely,
\begin{equation}
    P_\mathrm{vs,ZnS} = 10^{12.8117 - 15873\,\mathrm{K}/T - \left[\mathrm{Fe/H}\right]} \,\,\mathrm{bar}\,\, .
\end{equation}
As \ce{KCl} cloud particles may serve as condensation nuclei for \ce{ZnS} clouds,  considering them as separate cloud particles as done here may not be appropriate. This is not accounted for in the {\sc EddySed} model as implemented but has been investigated by \citet{gao_2018} using the {\sc CARMA} microphysics code.

 To couple {\sc EddySed} radiatively to the atmosphere, at each point in the atmosphere and for each cloud species the particle size distribution is resolved into 54 size bins between the sizes of $10^{-9}\,\mathrm{m}$ to $1.1\times 10^{-3}\,\mathrm{m}$ , allowing for the radiative properties (opacity, single-scattering albedo, and asymmetry parameter) to be computed using pre-calculated tables.  The cloud distribution and radiative properties are computed every radiative timestep (see Table \ref{Tbl:GasProp}). As the coupling of {\sc EddySed} to the atmosphere is done in such a way that the advection of condensate is not considered, the thermal impact of evaporation and condensation is neglected. Latent heat release was included in the model of \citet{charnay_2015b} and was found not to impact the dynamics. A more general discussion of the impact of latent heating in brown dwarf and exoplanetary atmospheres can be found in \citet{tan_2017}.

\subsection{Initial Conditions and Parameter Study}
\label{Sec:IC}
To initialise our simulations we use one-dimensional pressure-temperature profiles for GJ~1214b generated using {\sc Atmo} to construct a three-dimensional atmosphere initially in hydrostatic equilibrium and without any winds.  Simulations are then run for 600 days without clouds in order to spin-up the atmosphere at which point clouds are enabled.  Simulations are then run for an additional $400$ days with clouds enabled, except for the $100\times$ solar metallicity $f_\mathrm{sed}=0.5$ and $0.1$ cases.  These two cases are run for $800$ days with clouds enabled due to the greater impact of clouds on the atmosphere. Clear sky simulations are run for 1000 days to allow for an appropriate comparison.

For our main parameter study, we investigate two atmospheric metallicities, solar and $100\times$ solar, and three values of the sedimentation factor $f_\mathrm{sed}$ for the {\sc EddySed} cloud models: $f_\mathrm{sed}=1.0$, $0.5$, and $0.1$.  These choices of sedimentation factor are informed by previous one-dimensional models of clouds on GJ~1214b which found that clouds need significant vertical extent to match observations \citep{morley_2013}.    

To avoid issues associated with the sudden introduction of a new source of opacity, cloud opacities are linearly increased from zero to their full values over the course of the first 100 days after clouds are enabled (i.e., between days 600 and 700).  

\section{Results}
\label{Sec:Results}
In this section, we present the results of our simulations, first focusing on the impact of cloud on the atmospheric structure followed by an investigation of the impact of clouds on the synthetic observations.  A discussion of the convergence of the simulations is found in Appendix \ref{Apdx:Conv}.

\subsection{Atmospheric Structure }

\begin{figure*}
	\includegraphics[]{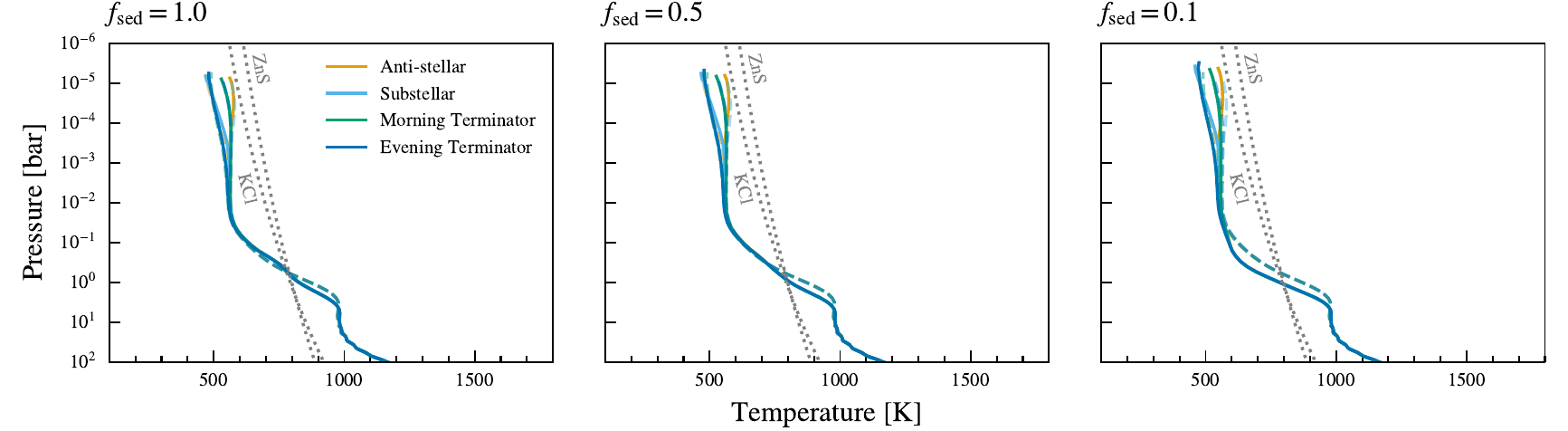}
    \caption{Pressure-temperature profiles for the case of a solar metallicity atmosphere. The cloudy cases are shown as solid lines while the clear-sky case is shown as a dashed line, for comparison. Dotted grey lines indicate the condensation curves for \ce{KCl} and \ce{ZnS}. }
    \label{Fig:PT_1x}
\end{figure*}

\begin{figure*}
	\includegraphics[]{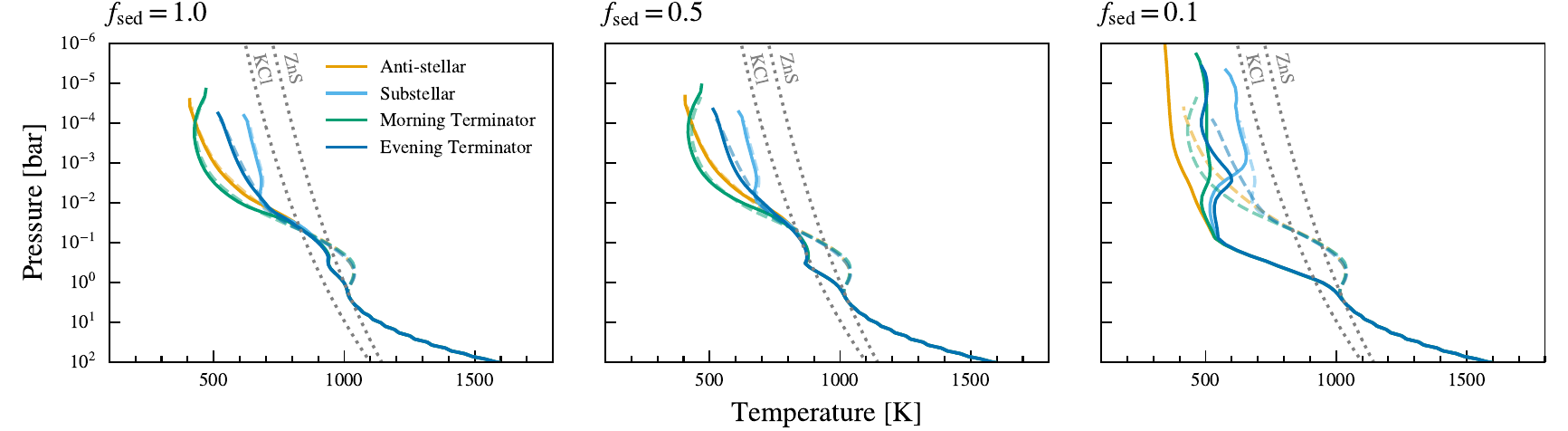}
    \caption{Pressure-temperature profiles for the case of a $100\times$ solar metallicity atmosphere. The cloudy cases are shown as solid lines while the clear-sky case is shown as a dashed line, for comparison. Dotted grey lines indicate the condensation curves for \ce{KCl} and \ce{ZnS}.}
    \label{Fig:PT_100x}
\end{figure*}

\subsubsection{Thermal Structure and The Distribution of Clouds}
\label{Sec:Thermal}

\begin{figure*}
	\includegraphics[]{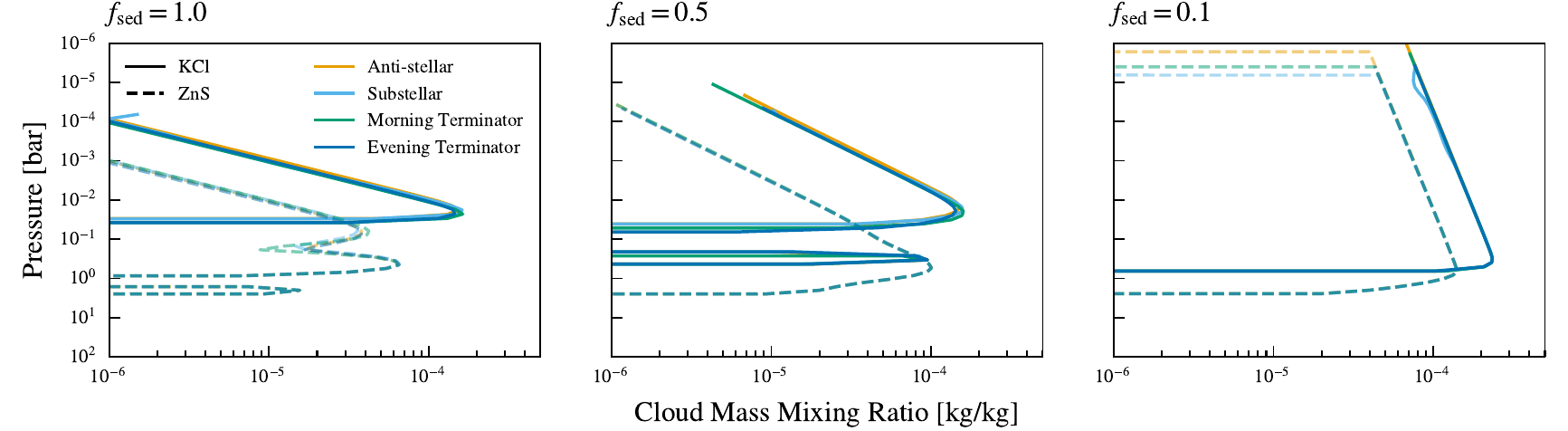}
    \caption{Cloud mass mixing ratio profiles for the case of a $100\times$ solar metallicity atmosphere. \ce{KCl} clouds are shown as solid lines while \ce{ZnS} clouds are dashed lines. }
    \label{Fig:cloud_100x}
\end{figure*}

The equatorial temperature profiles for the solar and $100\times$ solar metallicities are shown in Figures \ref{Fig:PT_1x} and \ref{Fig:PT_100x}, respectively. In each panel, the solid lines correspond to cloudy simulations and the dashed line is the clear sky case for the appropriate metallicity. In the case of solar metallicity (Figure \ref{Fig:PT_1x}), the introduction of cloud results in only a small ($\sim 50 \,\mathrm{K} $) shift to cooler temperatures between $0.1$ and $1.0$ bar. Due to the relatively small changes in the temperature structure, in the phase equilibrium limit the distribution of clouds resemble the case of post-processing a clear-sky simulation (not shown) with \ce{ZnS} clouds forming at $1\,\mathrm{bar}$ and \ce{KCl} forming at $0.1\,\mathrm{bar}$. The existence of \ce{ZnS} clouds below the \ce{KCl} cloud deck depends strongly on the existence of nucleation sites, either of \ce{KCl}, haze, or another form of dust grain \citep{gao_2018}, and as such the presence of \ce{ZnS} clouds at these pressures should be viewed with an increased level of uncertainty.   
In the case of $100\times$ solar metallicity (Figure \ref{Fig:PT_100x}), the addition of clouds has an increased impact over the lower metallicity case due to the scattering of stellar radiation back into space by \ce{KCl} clouds, reducing the net heating. While in the clear sky case only 0.8\% of the incoming stellar radiation is scattered back through the top of the atmosphere, the cloudy simulations show 7.8\%, 15.9\% and 44.9\% of incoming stellar radiation scattered back for $f_\mathrm{sed}=1.0$, $0.5$ and $0.1$, respectively.  In the $f_\mathrm{sed}=1.0$ case where clouds are relatively shallow, this results in a shift of $\sim 100\,\mathrm{K}$ near the \ce{ZnS} cloud deck at $1\,\mathrm{bar}$. As the clouds increase in vertical extent (i.e., with decreasing $f_\mathrm{sed}$), the impact of the reduced heating extends throughout the atmosphere at pressures below $1\,\mathrm{bar}$, with up to a $\sim 300\,\mathrm{K}$ shift around $P\sim 0.1\,\mathrm{bar}$ for the $f_\mathrm{sed}=0.1$ case. This shift in temperature results in the base of the clouds appearing at higher pressures (see Figure \ref{Fig:cloud_100x}).

Above the cloud deck, the vertical distribution of cloud exhibits the pressure dependence $q_\mathrm{c}\propto p^{f_\mathrm{sed}}$ characteristic of the {\sc EddySed} cloud model.  Horizontally, however, there is not significant variation in the cloud mixing ratio.  This is in part due to the fact that the although horizontal variation in temperatures are seen the cloudy simulations (see Fig. \ref{Fig:PT_100x}), the temperatures never approach or cross the condensation curve.  The lack of observed horizontal variation in cloud is likely also due, in part, to limitations of the modeling approach. A setup that more directly couples the vapour abundance to the local transport processes, instead of assuming a global average mixing rate, has the potential to show more horizontal variation (see, for example, Figure 2 in \citealt{charnay_2015b}).

\begin{figure*}
	\includegraphics[]{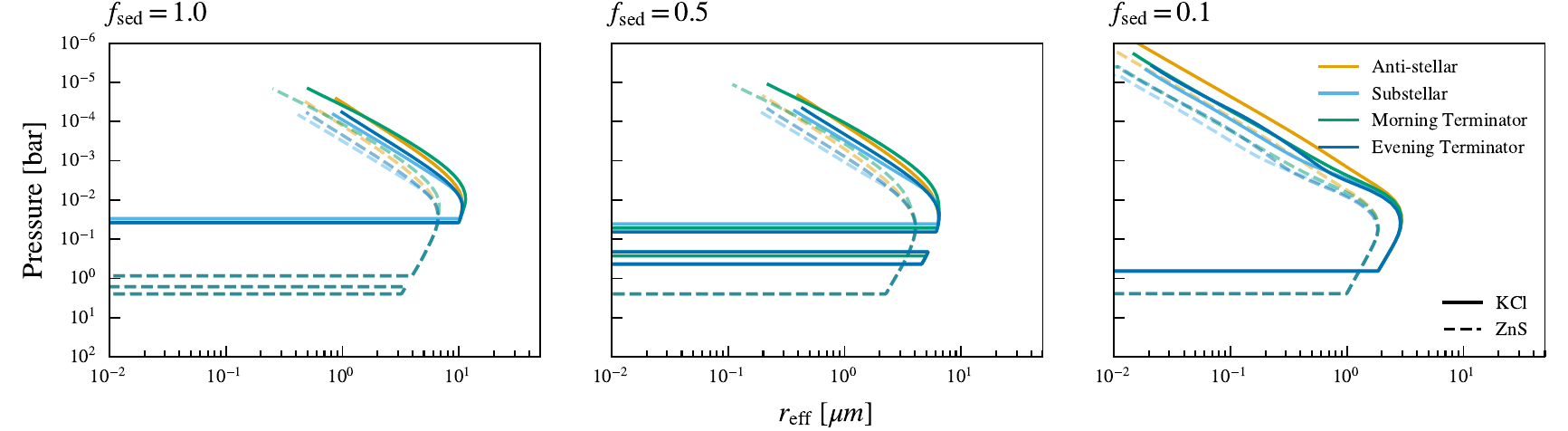}
    \caption{Profiles of the effective radius $r_\mathrm{eff}$ for the case of a $100\times$ solar metallicity atmosphere. KCl clouds are shown as solid lines while ZnS clouds are dashed lines.   }
    \label{Fig:cloud_rg_100x}
\end{figure*}

The effective particle sizes within the clouds (see Eqn. \ref{Eqn:reff} and Fig. \ref{Fig:cloud_rg_100x}) decrease with increased cloud extent as, by assumption, the reduced rainout necessitates smaller radii for the particles to remain suspended. The variation of particle sizes with pressure takes on a form characteristic of {\sc EddySed} where initially particle sizes increase with decreasing pressure due to the increase in vertical mixing but eventually turn over and begin to decrease with pressure as the atmosphere becomes increasingly more inefficient at supporting cloud particles. The effective particle sizes peak around pressures of $10^{-2}$ bar with sizes on the order of $1$ to $10$ microns. As the particle sizes in {\sc EddySed} are determined solely by the local conditions, the particle sizes become small as atmospheric pressures decrease. These small particles in the upper atmosphere are responsible for the scattering of the incident stellar radiation.

\subsubsection{Velocity Structures}

\begin{figure*}
	\includegraphics[]{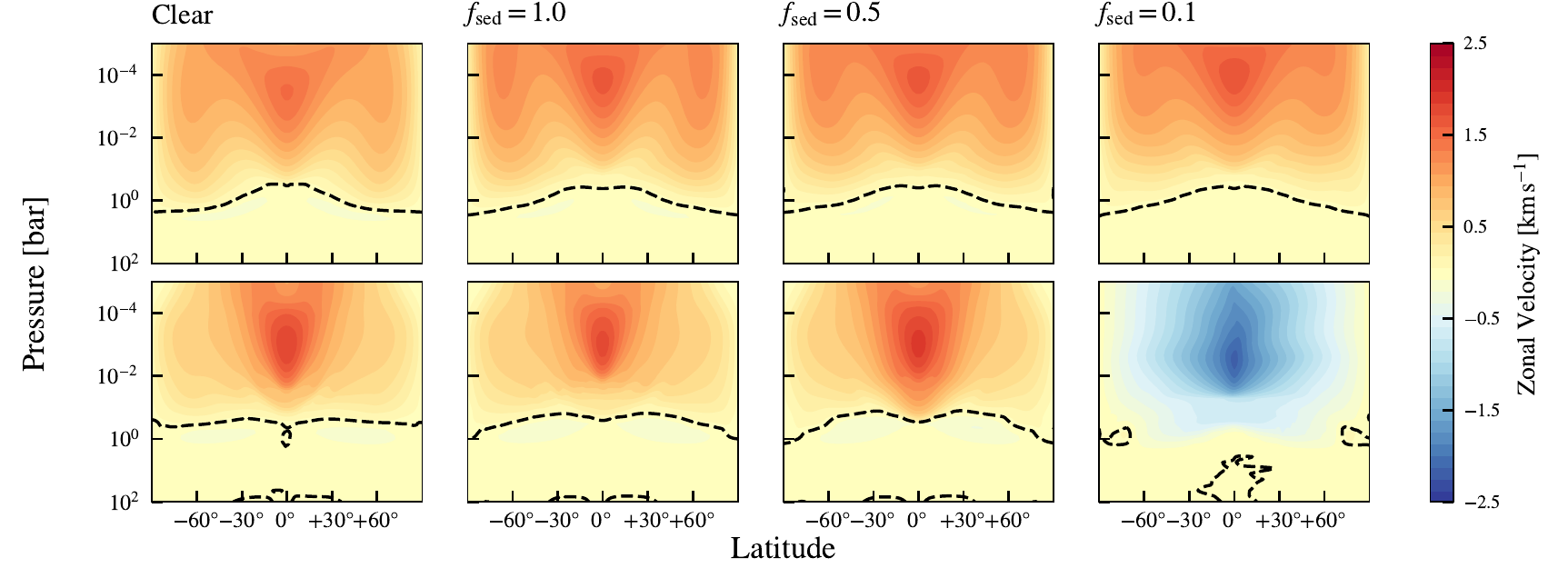}
    \caption{Zonal velocities for the solar metallicity case (top row) and the $100\times $ solar metallicity case (bottom row). Positive values indicate superrotation. The black dashed lines indicate the contour of zero zonal velocity. }
    \label{Fig:zonaljet}
\end{figure*}

There has been a lot of investigation of the zonal velocities in GCM simulations of GJ~1214b and the inconsistencies across simulations, although the differences in heating profiles between temperature forcing \citep{mayne_2019}, dual band grey radiative transfer \citep{menou_2012,wang_2020}, and multi-band correlated-k radiative transfer \citep{kataria_2014,charnay_2015a,charnay_2015b} at least in part explains the differences\footnote{Also relevant is the investigation of the impact of choice of radiative transfer scheme on models of the hot Jupiter HD~209458b by \citet{lee_2021}.  Relevant to the discussion here, they found differing temperature structures in the upper atmosphere and differing widths of the central jet between their dual grey and correlated-k models.}.  
In the simulations presented here, we find an equatorial jet forms in all cases except the $100\times$ solar metallicity, $f_\mathrm{sed}=0.1$ case (see Figure \ref{Fig:zonaljet}).  For the solar metallicity cases, we find two additional polar jets with amplitudes $\sim 1.2$ to $1.3\,\mathrm{km\, s^{-1}}$, qualitatively consistent with the results of \citet{charnay_2015a}, with limited impact from the introduction of clouds. The speed of the of the equatorial jet also remains consistent at $\sim 1.6\,\mathrm{km\, s^{-1}}$.     

For atmospheres with $100\times$ solar metallicity, we find superrotation in the mid-latitude and polar regions in the clear sky, $f_\mathrm{sed}=1.0$, and $f_\mathrm{sed}=0.5$ cases; however, there does not exist a jet separate from the equatorial jet.  Since the models do not include any drag at the lower boundary, a counterrotating flow also forms below the jets, roughly at pressures of $0.1$ to $1$ bar as angular momentum is carried upwards into the jet (see Figure \ref{Fig:zonaljet} and \citealt{showman_2011a}). The speed of the equatorial jet does appear to increase with cloud scale height; however, in the $f_\mathrm{sed}=0.1$ case, the superrotating jet eventually decays, forming two dayside gyres and a counter-rotating wind (see Figure \ref{Fig:horizvel} as well as additional plots in Appendix \ref{Apdx:Plots}). The mechanism causing the decay of the jet is unclear, although it does appear to be related to the increased opacity and the contracted temperature profile. To explore this, we ran two additional simulations (not shown) with $100\times$ solar metallicity and $f_\mathrm{sed}=0.1$ with the cloud opacity introduced at the start and no opacity ramp included. In the first test, we used the same initial temperature pressure profile outlined in Section \ref{Sec:IC} while in the second we used a cooler profile approximating the pressure-temperature profile in the end state of our $f_\mathrm{sed}=0.1$ case shown in Figure \ref{Fig:PT_100x}.  In the former case, the atmosphere spins up, forming an equatorial jet, before undergoing the same decay seen in our main simulations.   In the latter case, no transient equatorial jet forms, with the final state being qualitatively the same as in \ref{Fig:horizvel}, right panel.  We leave a more detailed investigation of the underlying mechanism and whether it is numerical in origin for future work.

\begin{figure*}
	\includegraphics[]{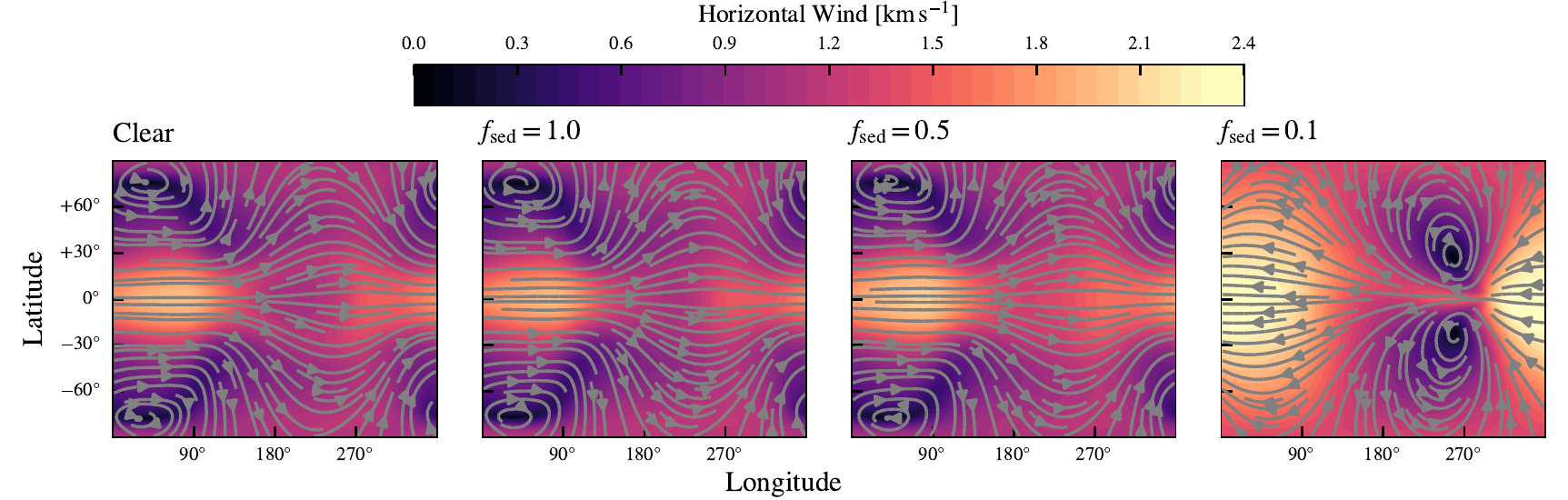}
    \caption{The horizontal wind speed at $0.001\,\mathrm{bar}$ for each of the $100\times$ solar metallicity cases. The horizontal winds show limited variation between cases except for the $f_\mathrm{sed}=0.1$ case where the nightside gyres move closer to the equator resulting in flow at the mid-latitudes on the nightside moving in a counter-rotating direction. }
    \label{Fig:horizvel}
\end{figure*}

\subsection{Observational Diagnostics}
To allow for comparison with observations, we have generated synthetic observations using the {\sc Socrates} radiative transfer routines within the {\sc UM} with the spectral resolution increased to $10\, \mathrm{cm^{-1}}$. This approach allows for the synthetic observations to be generated self-consistently using the same routines and opacity data used in the heating and cooling calculations. These routines have previously been used in \citet{lines_2018b,lines_2019,christie_2021} and the specifics of the transmission spectrum calculation are presented in \citet{lines_2018b}.

\subsubsection{Transmission Spectra}
We calculate transmission spectra for each of the simulations as in \citet{christie_2021}, with the exception that the model results are not scaled to agree with the observed value at $1.4$ microns, as was the case in \citet{christie_2021}. As the results from the solar metallicity simulations show little impact from the introduction of clouds, we opt to focus on the results from the $100\times$ solar metallicity simulations here.

\begin{figure*}
	\includegraphics[]{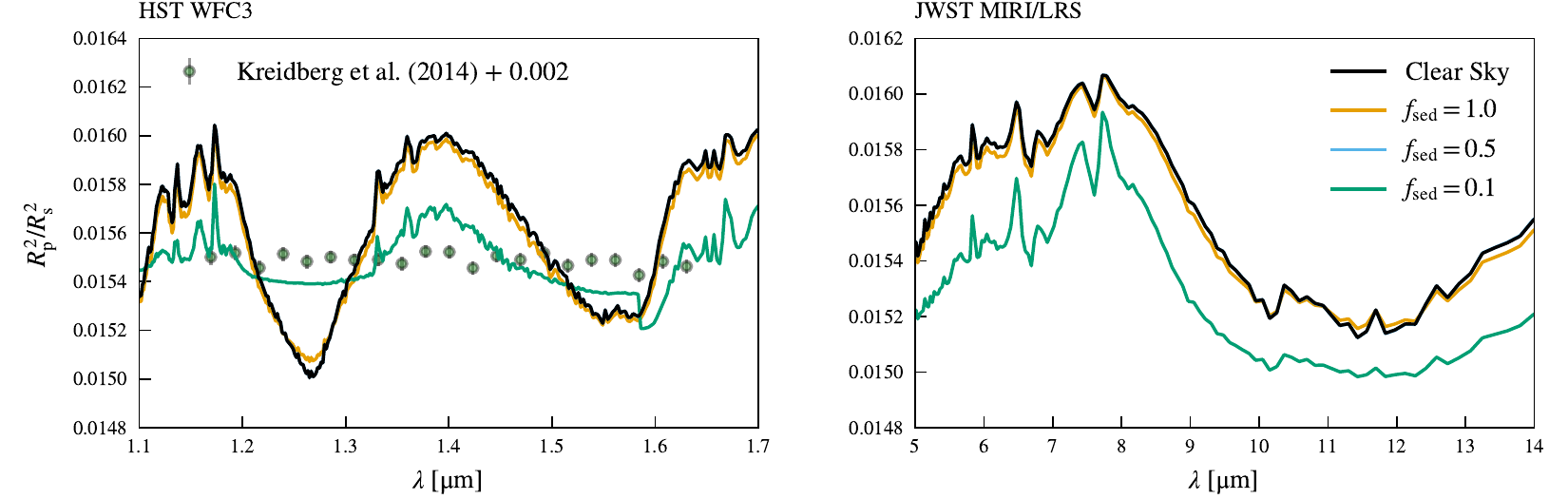}
    \caption{The simulated transmission spectra for the HST WFC3 (left) and JWST MIRI/LRS (right) bandpasses for the $100\times$ solar metallicity simulations.  The observational data from \citet{kreidberg_2014} are also included in green, and the data have been shifted upwards by $0.002$ to facilitate comparison.  }
    \label{Fig:transmission}
\end{figure*}

The left panel of Figure \ref{Fig:transmission} shows the synthetic observations for the HST WFC3 bandpass covering $1.1$ to $1.7\, \mathrm{\mu m}$.  The observations of \citet{kreidberg_2014} are included, with the data shifted up by $0.002$ to facilitate comparison.\footnote{As the appropriate radius for the $200\,\mathrm{bar}$ inner boundary is unknown, a uniform shift in the transmission spectrum approximates a move in the location of the inner boundary. To properly address this issue, the appropriate value of the inner boundary radius could be found by searching the parameter space of possible radii; however, this would be computationally expensive for what is likely little actual gain. } A shallow cloud layer, as in the $f_\mathrm{sed}=1.0$ and $f_\mathrm{sed}=0.5$ cases, has only a minor impact on the transmission spectrum. The $f_\mathrm{sed}=0.1$ case shows the best agreement with the data, insofar as the spectrum is relatively flat, however, the $1.4$ micron \ce{CH_4} feature is still visible. The sharp decrease in the transmission spectrum of the $f_\mathrm{sed}=0.1$ case near $1.6$ microns is due to insufficient precision in the fixed-point recording of the refraction index data for \ce{ZnS} and is unlikely to be physical in origin \citep[see][]{querry_1987}.

\begin{figure*}
	\includegraphics[]{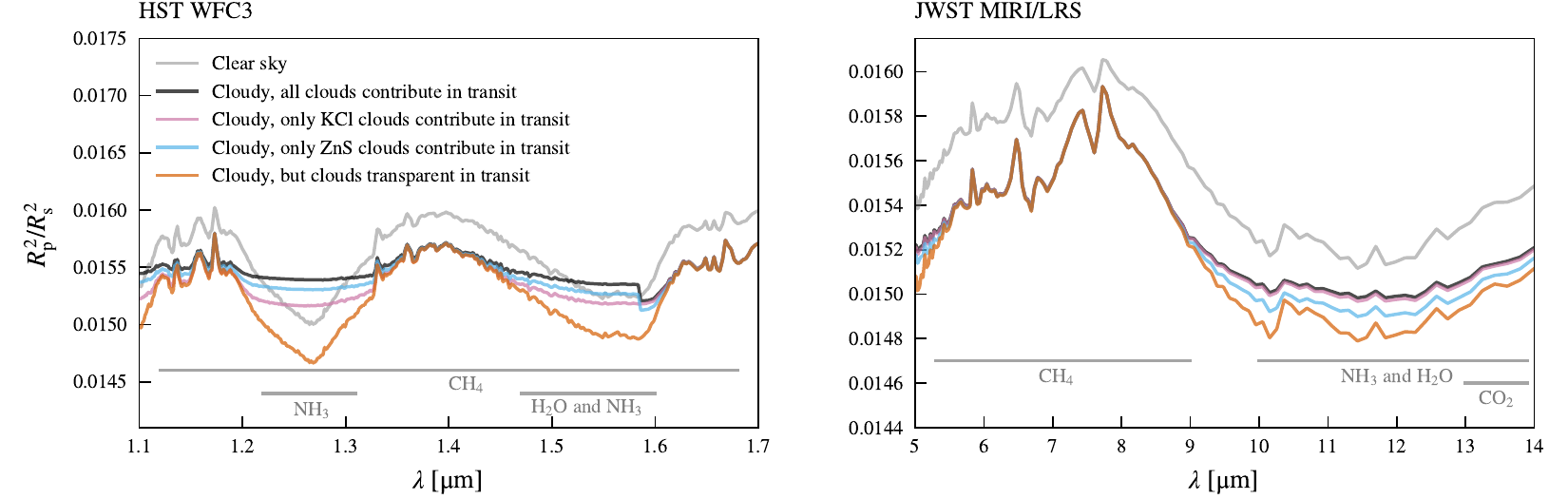}
    \caption{The simulated transmission spectra for the HST WFC3 (left) and JWST MIRI/LRS (right) bandpasses for the $100\times$ solar metallicity clear sky case and the cloudy $100\times$ solar metallicity $f_\mathrm{sed}=0.1$ case allowing for  differing contributions to the transmission spectrum for each of the cloud species.  The grey lines at the bottom show the major gas-phase contributors in the indicated parts of the spectrum. }
    \label{Fig:transbd}
\end{figure*}

To better understand how the specific cloud species are impacting the transmission spectrum in the $f_\mathrm{sed}=0.1$ case, we breakdown the spectrum further. Figure \ref{Fig:transbd} shows the transmission spectrum along with post-processed versions of the same run where different combinations of cloud species contribute to the opacity in the transmission spectrum calculation.  The transmission spectrum for the clear sky case is shown in grey for comparison. The impact of the clouds on the thermal profile and the extent of the atmosphere as it contracts thermally can be seen in a relatively uniform shift to smaller values in the transmission spectrum from the true clear sky spectrum to the cloudy spectrum where clouds contribute to the heating and cooling but are transparent in transit. Looking instead at the individual contributions to the transmission spectrum, the largest contribution can be seen to be \ce{ZnS}. As we have made assumptions favourable to the formation of \ce{ZnS} clouds in that there are always sufficient condensation nuclei and there are no energy barriers limiting condensation (see \citealt{gao_2018} for a full discussion), the models may be overestimating the impact of \ce{ZnS}.

We additionally look at the transmission spectra in the JWST MIRI/LRS bandpass (see Figure \ref{Fig:transmission}, right panel), motivated by the possibility of future observations \citep{bean_2021}.  As in the previous analysis of the WFC3 bandpass, clouds show the largest impact in the $f_\mathrm{sed}=0.1$ case.  The cooler atmosphere results in a smaller transit across all wavelengths in the bandpass (see Figure \ref{Fig:transbd}, right panel).  In the region between  $6\,\mathrm{\mu m}$ and $9\,\mathrm{\mu m}$, the spectrum is dominated by \ce{CH_4} without any direct impact from clouds seen in our synthetic spectrum.  At longer wavelengths, however, \ce{KCl} clouds,  and to a lesser extent \ce{ZnS} clouds, flatten the spectrum, obscuring spectral features (see, for example, the \ce{NH_3} feature between $10\,\mathrm{\mu m}$ and $11\,\mathrm{\mu m}$ in Figure \ref{Fig:transbd}, right panel). The limited impact of \ce{ZnS} in this bandpass may be in part to the limitations in the \citet{querry_1987} refraction index data, discussed above. 

\subsubsection{Phase Curves}
\begin{figure*}
	\includegraphics[]{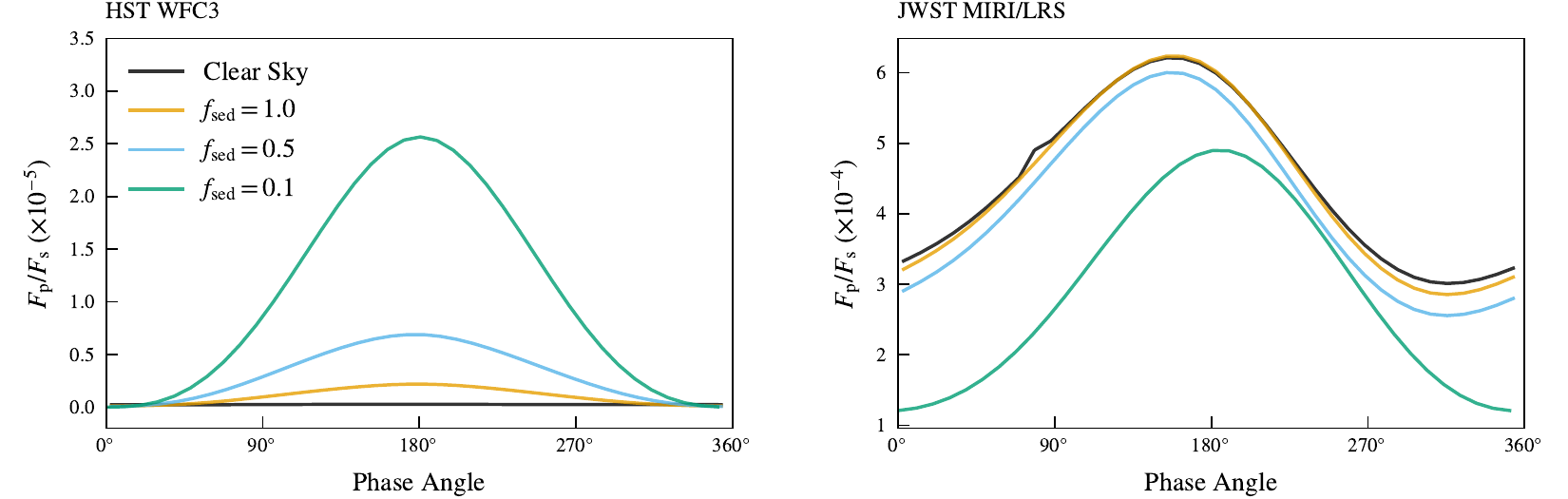}
    \caption{Simulated phase curves for each of the $100\times$ solar metallicity simulations for the HST WFC3 bandpass (left, with a wavelength range between $1.1$ and $1.7\,\mathrm{\mu m}$) and the JWST MIRI/LRS bandpass (right, with a wavelength range between $5$ microns and $12\,\mathrm{\mu m}$).  }
    \label{Fig:phase}
\end{figure*}

Although there do not currently exist phase curve observations of GJ~1214b, there are planned observations using JWST MIRI/LRS \citep{bean_2021}. These observations cover the wavelength range between $5$ and $12$ microns and are expected to be an excellent probe of the thermal structure of the planet due to the minimal impact of scattering by aerosols. Figure \ref{Fig:phase}  (right panel) shows the synthetic phase curves for each simulation with $100\times$ solar metallicity. As expected given the small impact of the cloud on the thermal structure in the $f_\mathrm{sed}=1.0$ case, there is minimal difference between the clear sky and $f_\mathrm{sed}=1.0$ phase curves. The simulation with the most vertically extended clouds -- the $f_\mathrm{sed}=0.1$ case-- shows a decrease in the peak of the phase curve associated with the cooler atmospheres which can be contrasted with previous cloudy models of the hot Jupiter HD~209458b where dayside temperatures increased with cloud scale height (i.e., with decreasing $f_\mathrm{sed}$) resulting in higher peaks in phase curves \citep{christie_2021}.  Despite these differences, the introduction of cloud still results in an increase in contrast in the phase curves between the minima and maxima (a factor of $2.1$ for the clear case compared to $4.0$ for the $f_\mathrm{sed}=0.1$ case) and a shift of the peak back towards $180$\textdegree, as discussed in \citet{parmentier_2020} and seen in our previous work \citep{christie_2021}. The lack of offset is also likely due to the lack of equatorial jet pushing the hot spot westward. We also note that, as expected for the given wavelength range, the phase curve is probing thermal emission with only $\sim 10^{-9}\%$ of the emission coming from reflected starlight in the clear sky case and increasing to $\sim 2\%$ in the $f_\mathrm{sed}=0.1$ case.

As the effective temperature of GJ~1214 is $3026\,\mathrm{K}$ \citep{charbonneau_2009}, the stellar blackbody peaks at $\sim 1\,\mathrm{\mu m}$ allowing for the possibility of scattered starlight by clouds contributing significantly to the HST WFC3 phase curve (see Figure \ref{Fig:phase}, left panel).  For the clear sky, $100\times$ solar metallicity case where there is limited scattering of starlight, the planetary thermal emission dominates the phase curve, with $F_\mathrm{p}/F_\mathrm{s}\,\sim 2.3$ to $2.7\times 10^{-7}$. With increased cloud extent, however, the scattered stellar component begins to dominate the phase curve, and for $f_\mathrm{sed}=0.1$, the phase curve peaks with $F_\mathrm{p}/F_\mathrm{s} = 2.6\times 10^{-5}$, with 99.8\% of the planetary flux coming from the scattered stellar component.

\section{Discussion and Conclusions}
\label{Sec:Conclusions}
In this paper we have investigated the impact of clouds on the dynamics and observables of the warm Neptune GJ~1214b, specifically through the coupling the one-dimensional phase equilibrium {\sc EddySed} cloud model to the UM GCM. Consistent with previous investigations, we find that increased metallicity and increased cloud vertical extent (i.e., decreased $f_\mathrm{sed}$) are necessary to impact the pressure temperature profiles and synthetic observables. The most significant impact was for our $100\times$ solar metallicity case with $f_\mathrm{sed}=0.1$ where $\sim 45\%$ of the incident stellar radiation was scattered back into space resulting in cooling and contraction of the atmosphere. These clouds, primarily the \ce{ZnS} component, also increased the atmospheric opacity within the spectral windows but did not entirely obscure the spectral features (see e.g., $1.4$ microns in Figure \ref{Fig:transmission}). To reproduce the flat spectrum of \citet{kreidberg_2014} the cloud content either needs to be further increased through alterations to the cloud model or the metallicity further increased, consistent with the conclusions of \citet{gao_2018}.   

Dynamically, the inclusion of clouds results in an increase in the speed of the equatorial jet, except in the $100\times$ solar metallicity case with $f_\mathrm{sed}=0.1$, where we see a suppression of the equatorial jet and the formation of two dayside gyres. It is unclear whether the latter configuration is a stable or whether it will continue to evolve, or to what extent it is numerical in origin. Understanding this better likely requires further improving our cloud models, and potentially moving to tracer-based models that better capture the localized dynamics and improved microphysics to properly model the particle size distribution.  The move to a tracer-based model would also facilitate the inclusion of photochemical hazes, which don't fit naturally in the {\sc EddySed} framework.

\section*{Acknowledgements}
We thank the anonymous referee for comments that greatly improved the quality of this paper.  The observational data were retrieved from Dr. Hannah Wakeford's online archive at www.stellarplanet.org. Material produced using Met Office Software. This research made use of the ISCA High Performance Computing Service at the University of Exeter. This work was performed using the DiRAC Data Intensive service at Leicester, operated by the University of Leicester IT Services, which forms part of the STFC DiRAC HPC Facility (www.dirac.ac.uk). The equipment was funded by BEIS capital funding via STFC capital grants ST/K000373/1 and ST/R002363/1 and STFC DiRAC Operations grant ST/R001014/1. DiRAC is part of the National e-Infrastructure. This work was partly funded by the Leverhulme Trust through a research project grant [RPG-2020-82], a Science and Technology Facilities Council Consolidated Grant [ST/R000395/1] and a a UKRI Future Leaders Fellowship [grant number MR/T040866/1]. For the purpose of open access, the authors have applied a Creative Commons Attribution (CC BY) licence to any Author Accepted Manuscript version arising.

\section*{Data Availability}
The research data supporting this publication are archived on the Harvard Dataverse and are available at \href{https://doi.org/10.7910/DVN/8XAEPY}{doi.org/10.7910/DVN/8XAEPY}.



\bibliographystyle{mnras}
\bibliography{references} 




\appendix

\section{Pseudo-Spherical Irradiation in the UM}
\label{Apdx:SphIrr}
In this work, we take advantage of the UM's implementation of pseudo-spherical irradiation \citep{jackson_2020} which calculates the attenuation of the incoming stellar radiation using spherical shells.  This is in contrast to the previous plane-parallel implementation where the attenuating column is approximated using the vertical column $N$, corrected using the zenith angle $\alpha$, yielding an effective attenuating column $N/\cos\alpha$. As a result, in the plane parallel implementation, shortwave heating goes to zero at the terminator, without any possibility of nightside heating, an issue remedied by the pseudo-spherical treatment. This treatment has also been applied to the study of the impact of flares on `Earth--like' terrestrial exoplanets in Ridgway et al. (submitted to MNRAS). In this section, we compare GJ~1214b clear sky results using the two different irradiation schemes to ascertain the impact of the change, and we demonstrate that for our test case the switch to pseudo-spherical irradiation results in slower jets near the poles.

\begin{figure*}
	\includegraphics[]{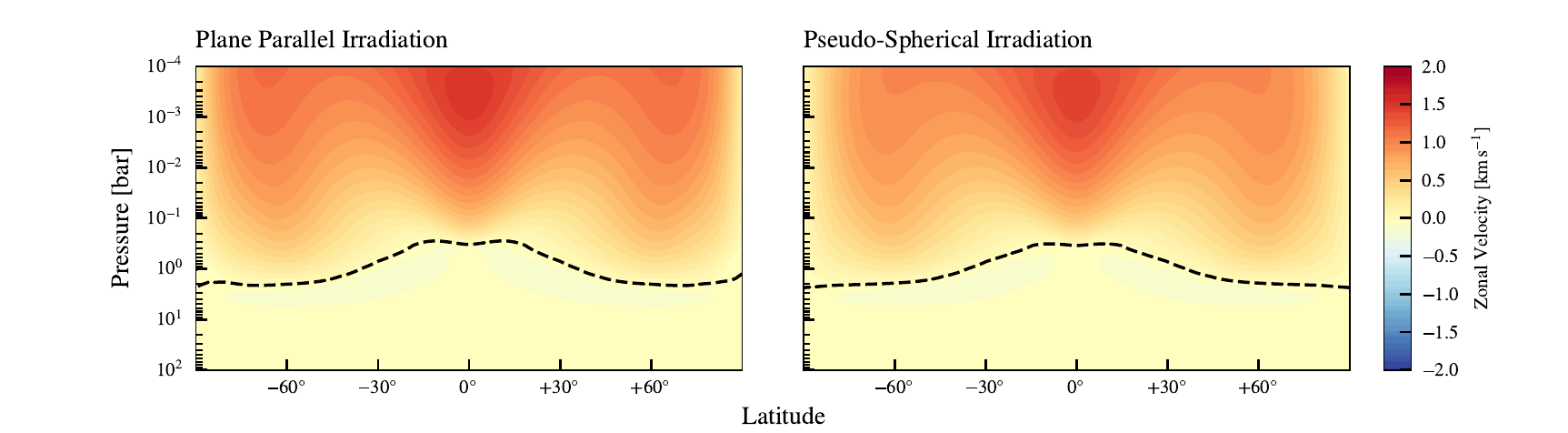}
    \caption{The zonally-averaged azimuthal velocity for a simulation with plane parallel irradiation (left) and   pseudo-spherical irradiation (right).  Both simulations are run without clouds for 1000 days.  The black dashed line indicates the contour of zero zonal velocity. }
    \label{Fig:SphIrrVel}
\end{figure*}

\begin{figure*}
	\includegraphics[]{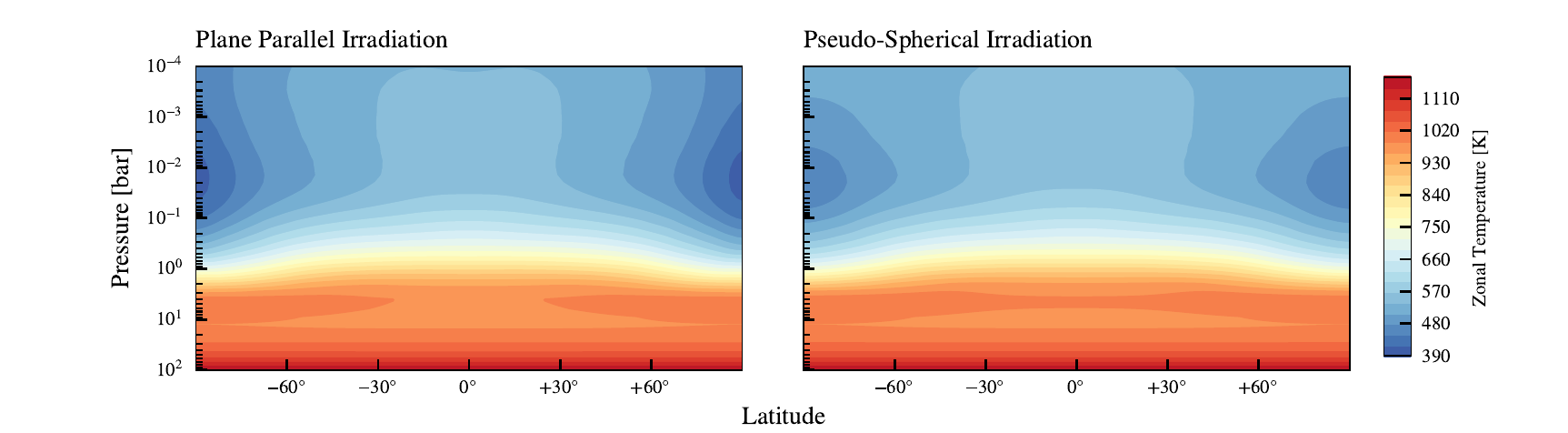}
    \caption{The zonally-averaged temperature for a simulation with plane parallel irradiation (left)  and  pseudo-spherical irradiation (right).  Both simulations are run without clouds for 1000 days.  }
    \label{Fig:SphIrrTemp}
\end{figure*}

\begin{figure*}
	\includegraphics[]{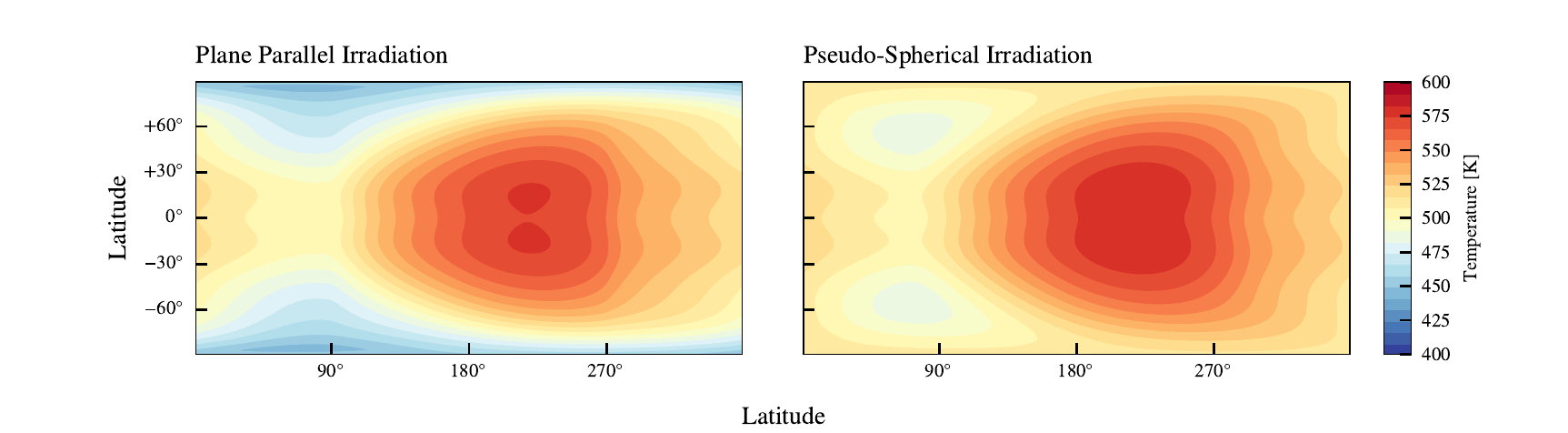}
    \caption{The temperature at $10^{-4}\, \mathrm{bar}$ for a simulation with plane parallel irradiation (left) and pseudo-spherical irradiation (right).  Both simulations are run without clouds for 1000 days. }
    \label{Fig:SphIrrTemp2}
\end{figure*}

For this test, we ran each of the two cases for 1000 days using the clear-sky solar-metallicity configuration outlined in the paper. Qualitatively, we see the same velocity structures forming in both cases (see Figure \ref{Fig:SphIrrVel}), with a fast equatorial jet and two polar jets with the equatorial jet having a maximum speed of $\sim 1.5\,\mathrm{km\,s^{-1}}$. The impact of the new irradiation scheme is primarily felt at the poles where shortwave radiation heating the nightside reduces the temperature contrast around the poles resulting in a slower polar jet ($1.2\, \mathrm{km\,s^{-1}}$ in the plane-parallel case versus $1.0    \,\mathrm{km\,s^{-1}}$ in the spherical case). This can be seen in Figures \ref{Fig:SphIrrTemp} and \ref{Fig:SphIrrTemp2} where the plane-parallel case shows a larger decrease in the temperature approaching the pole. 

We estimate the energy deposited in the atmosphere based on the heating rates and find that the spherical irradiation scheme results in $\sim 6\%$ more atmospheric heating than the plane parallel case.   While in theory the atmospheric heating should be insensitive to the radiation scheme, losses do occur due to various assumptions being made, and we will investigate this further in a follow-up paper.

We note that while using spherical irradiation makes modest changes to the structure of the atmosphere in the regions of interest for studies like this one, it does better characterize the underlying radiative transfer within the atmosphere, and it may offer some improvement in stability as we see a reduction in both horizontal and vertical velocities near the poles.

\section{{\sc EddySed} with a Gamma Distribution}
\label{Apdx:Gamma}
Without direct observations of clouds on exoplanets, it is difficult to constrain the appropriate particle size distribution to use in the modelling of clouds on these planets; however, in water clouds on Earth, where in-situ observations of raindrop and snowflake sizes are possible, the size distributions are often fit to mono- or multi-modal log-normal \citep[e.g.,][]{ackerman_2000}, exponential \citep[e.g.,][]{marshall_1948}, or gamma distributions \citep[e.g.,][]{ulbrich_1983}. Informed by this, models of exoplanetary clouds that do not explicitly track the particle sizes often assume one of these distributions \citep[e.g.,][]{ackerman_2001,helling_2008,charnay_2018}. 


In this appendix, we look at the impact of switching from a log-normal particle size distribution to a gamma particle size distribution within the {\sc EddySed} framework. While the assumption of a log-normal particle size distribution is often the simplest from a computational and analytical standpoint, {\sc EddySed} can be formulated using a gamma distribution\footnote{The gamma distribution is someitmes presented as the {\em Hansen distribution} \citep[e.g.,][]{hansen_1971,burningham_2021} or the {\em potential exponential distribution} \citep[e.g.,][]{helling_2008} in different forms in the literature.} which is presented below. 


\subsection{Derivation}
Considering a layer of cloud within the atmosphere, we take the number distribution of cloud particles to be $dn/dr = N f(r; A,B)$ where N is the total number of particles and $f(r;A,B)$ is the gamma distribution, 
\begin{equation}
f(r; A,B) = \frac{B^Ar^{A-1}}{\Gamma(A)}\exp\left(-Br\right).
\end{equation}
\noindent The two parameters A and B are more often written as $\alpha$ and $\beta$; however, as those are used elsewhere in paper, we opt for this simple substitution in nomenclature. In \citet{ackerman_2001}, the variance of the distribution is  $\Var\left[\log X\right] = \log^2\sigma$ where $\sigma$ parametrises the width of the distribution. For a gamma distribution, the variance of a gamma-distributed variable $X$ depends only on the parameter $A$, 
\begin{equation}
	\Var\left[\log X\right] = \psi^{(1)}(A),
\end{equation}
\noindent where $\psi^{(1)}(x)$ is the trigamma function.  Following the method of \citet{ackerman_2001}, we can then fix the value of $A$ based on a prescribed $\sigma$,
\begin{equation}
\psi^{(1)}(A) = \log^2\sigma\,\, .
\end{equation}
\noindent As $\sigma$ is taken to be constant throughout the atmosphere in the {\sc Eddysed} model, $A$ only needs to be calculated at the beginning of the simulation, limiting computational overhead. For the default value of $\sigma=2$ used here, $A \approx 2.54278$.  With $A$ fixed, we can now calculate $B$ based on the sedimentation arguments of \citet{ackerman_2001}. The sedimentation factor is defined as
\begin{equation}
f_\mathrm{sed} = \frac{\int_{0}^{\infty} r^{3+\alpha}(dn/dr)dr}{r_w^\alpha \int_{0}^{\infty} r^3 (dn/dr)dr}.
\end{equation}
\noindent This can be rewritten as
\begin{equation}
f_\mathrm{sed} = \frac{\E\left[r^{3+\alpha}\right]}{r_w^\alpha\E\left[r^3\right]} = \frac{1}{(Br_w)^\alpha}\frac{\Gamma(A+3+\alpha)}{\Gamma(A+3)},
\end{equation}
where $E[x]$ is the expected value of $x$.  We can then solve for $B$,
\begin{equation}
B = \frac{1}{r_w}\left[\frac{\Gamma(A+3+\alpha)}{f_\mathrm{sed}\Gamma(A+3)}\right]^{1/\alpha}.
\end{equation}
In the rare cases where $\alpha$ is an integer, the identity $\Gamma(z+1) = z\Gamma(z)$ can be used to eliminate the gamma function entirely and reduce the equation to
\begin{equation}
B = \frac{1}{r_w}\left[\frac{(A+2+\alpha)\cdots(A+3)}{f_\mathrm{sed}}\right]^{1/\alpha}\,\, .
\end{equation}
\noindent While this is worth noting, we will not use this simplification for the rest of the derivation as $\alpha$ is generally not an integer.

We can then compute the total number of particles based on mass conservation
\begin{equation}
q_\mathrm{c}\rho_a = \frac{4}{3}\pi N\rho_\mathrm{c}\int_{0}^{\infty}r^3 f(r;A,B)dr,
\end{equation}
\noindent resulting in 
\begin{equation}
N = \frac{3q_\mathrm{c}\rho_\mathrm{a}B^3}{4\pi \rho_\mathrm{c}(A+2)(A+1)A}.
\end{equation}

\begin{figure}
	\includegraphics[width=\columnwidth]{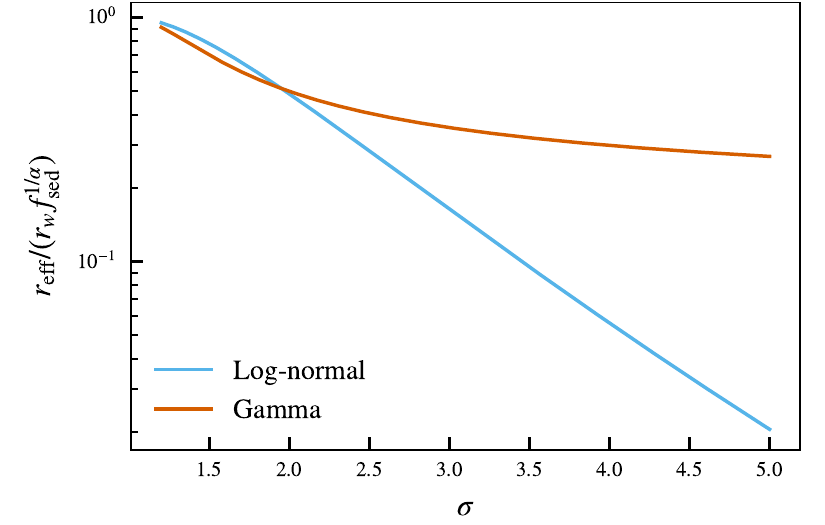}
    \caption{The effective particle radii $r_\mathrm{eff}$ for {\sc EddySed} formulated with a log-normal distribution (solid lines) and a gamma distribution (dashed lines).  For the purposes of this plot, $\alpha=2$ has been assumed.  }
    \label{Fig:reff_scaling}
\end{figure}

The area-weighted effective radius $r_\mathrm{eff}$ in a cloud layer is given by
\begin{equation}
r_\mathrm{eff} = \sqrt{\E[r^2]} = \left(\frac{A(A+1)}{B^2}\right)^{1/2}\,\, .
\end{equation}
\noindent  The scaling of $r_\mathrm{eff}$ with the size parameter $\sigma$ is very different in the case of a gamma distribution compared to the log-normal distribution. For small values of $\sigma$, the two distributions scale similarly with the distributions becoming sharply peaked with $r_\mathrm{eff} \sim r_wf_\mathrm{sed}^{1/\alpha}$; however, for wider distributions (e.g., for $\sigma$ greater than 2, see Figure \ref{Fig:reff_scaling}), $r_\mathrm{eff}$ decreases slower with $\sigma$ in the case of the gamma distribution case compared to the log-normal distribution case. For wider distributions, we may, as a result, expect larger differences in particle sizes and optical properties between cloud layers with a log-normal distribution than with cloud layers with a gamma distribution. Coincidentally, we find that for reasonable values of $\alpha$ the effective radii $r_\mathrm{eff}$ for the two distributions are similar for $\sigma \sim 2$, the value used here and widely in the literature.  

\begin{figure}
	\includegraphics[width=\columnwidth]{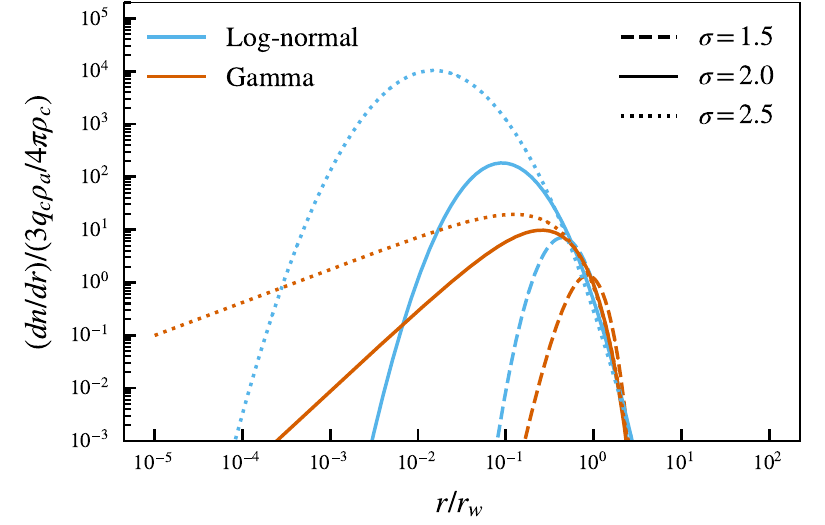}
    \caption{A comparison of the log-normal (red) and gamma (blue) distributions for three values of $\sigma$.  In each case, $f_\mathrm{sed}=1.0$ and $\alpha=2$ have been assumed. }
    \label{Fig:lognorm_v_gamma}
\end{figure}

To illustrate how using {\sc EddySed} with a gamma distribution differs from the default log-normal distribution case, the log-normal and gamma distributions are plotted in Figure \ref{Fig:lognorm_v_gamma} for $f_\mathrm{sed}=1.0$ and $\alpha=2$ and three values of $\sigma$. Compared to the log-normal distribution case, both location of the the peak of the gamma distribution and the number of particles show a much weaker dependence on $\sigma$ with the peak of the gamma distribution staying close to $r_w$ and the width being modulated by the low-mass tail. This weaker dependence and the skewness of the gamma distribution means that while varying $\sigma$ does change the particle numbers, the condensate mass remains more concentrated in the high mass end of the distribution compared to the log-normal distribution where a wider distribution results in more massive particles being traded for exponentially more particles at the peak of the distribution.

\subsection{Results}
\begin{figure}
	\includegraphics[]{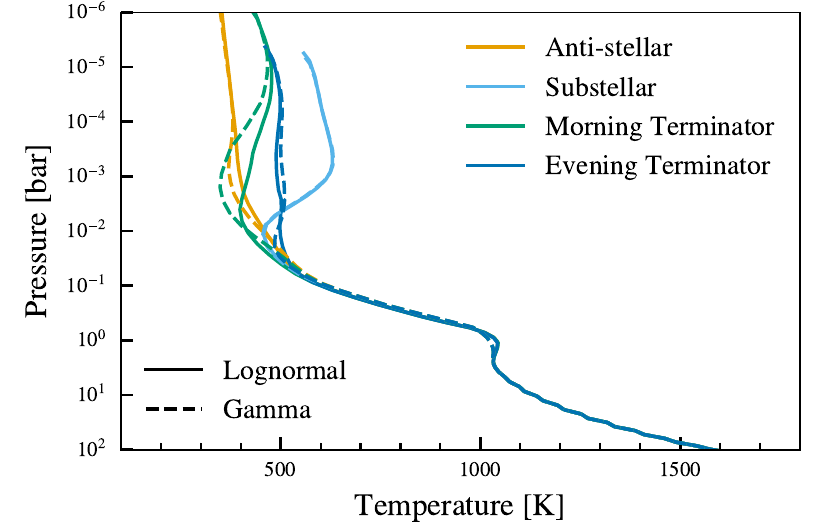}
    \caption{The temperature profiles at a latitude of $45^\circ$ for the $100\times$ metallicity, $f_\mathrm{sed}=0.1$ case for a log-normal distribution (solid lines) and a gamma distribution (dashed lines) assuming $\sigma=2$.   Equatorial profiles are not included as they did not differ noticeably between the two distributions.  }
    \label{Fig:SizeDist}
\end{figure}

\begin{figure*}
	\includegraphics[]{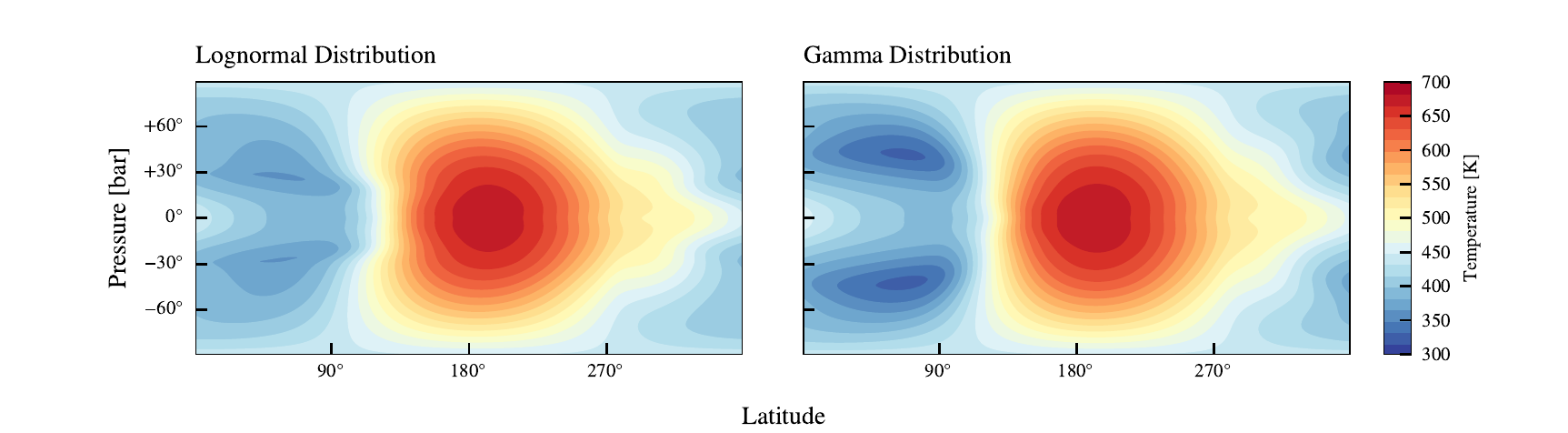}
    \caption{The temperatures at $1\,\mathrm{mbar}$ for the log-normal distribution case (left) and the gamma distribution case (right). }
    \label{Fig:SizeDist2}
\end{figure*}

To examine the impact of the change in distribution, we focus on the $100\times$ solar metallicity $f_\mathrm{sed}=0.1$ case as our analysis has already shown cloud in this case to have a noticeable impact, although we only compare the results after 1000 days.  Although larger values of $\sigma$ may result in larger differences between the two cases, we opt to keep $\sigma=2$ as it represents the standard value used here and in other studies. Comparing the two cases, we find negligible differences in the equatorial temperature structure; however, at mid-latitudes we find cooler temperatures on the night side and along the morning terminator in the case of a gamma distribution (see Figures \ref{Fig:SizeDist} and \ref{Fig:SizeDist2}). The differences occur around $P\sim 10^{-3}\,\mathrm{bar}$, with the log-normal distribution case being $\sim 100\,\mathrm{K}$ hotter than the gamma distribution case.  We note that the flow at mid-latitudes is somewhat different between the log-normal and gamma distribution cases, with the equatorial jet decaying at a slower rate in the case of the gamma distribution; however, longer run times are required to understand to what extent these differences are transient. This does not have a noticeable impact on the transmission spectrum, either through the change in atmospheric temperature or through any change in cloud opacity used in the transmission spectrum calculation.

Switching to a gamma distribution may result in larger differences from the log-normal distribution case in parts of the parameter space not investigated here (e.g., the case of very wide distributions).  We have opted to not investigate these cases as wide distributions quickly run into the issue of significant fractions of the particle distribution falling outside of the permitted or physically-plausible range of particle sizes at low pressures.  Relaxing or modifying the {\sc EddySed} assumption of a fixed distribution width may allow this issue to be avoided, but also it may provide an avenue for moving beyond {\sc EddySed} more generally.

\section{Comparison of Condensation Curves with {\sc ATMO}}
\label{Apdx:Cond}
\begin{figure}
	\includegraphics[width=\columnwidth]{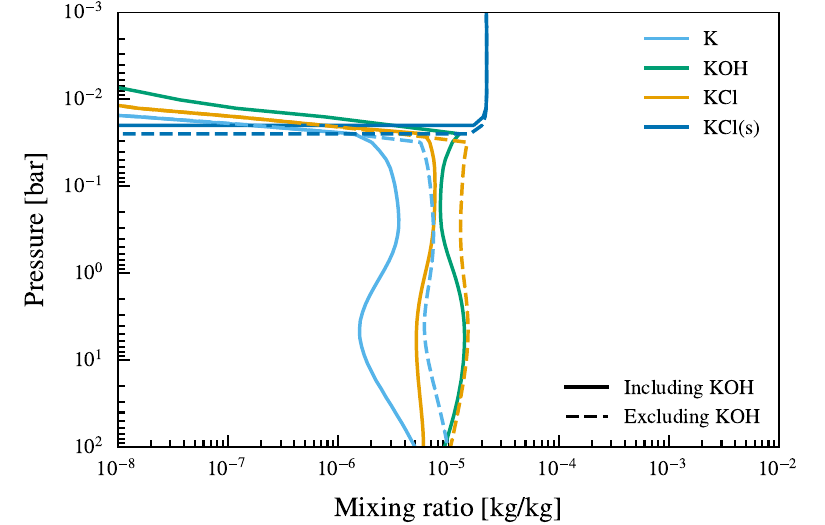}
    \caption{The chemical equilibrium for the major K-bearing species for the $100\times$ solar case, generated using the {\sc Atmo} code.  In one case, \ce{KOH} has been included (solid line) while in the other \ce{KOH} has been removed from the possible species (dashed line).  Although when \ce{KOH} is allowed to form it becomes the dominant K-bearing species below the \ce{KCl} cloud deck, it has a negligible impact on the formation of \ce{KCl(s)}.   }
    \label{Fig:KOH}
\end{figure}

\ce{KCl} clouds form directly from the condensation of gaseous \ce{KCl} which, at solar metallicity, is the dominant K-bearing species in the atmosphere. As metallicity increases, however, the abundance of \ce{KOH} increases, eventually exceeding \ce{KCl} in abundance.  To ascertain what impact this may have, we use the {\sc Atmo} chemistry code to compute the chemical equilibria for the initial $100\times $ solar temperature profile for two cases. In the first, we allow \ce{KOH} to form, but in the second, we artificially exclude \ce{KOH}.   When \ce{KOH} is allowed to form, there is a reduction in gaseous \ce{KCl}, resulting in it becoming supersaturated at a slightly lower pressure (see Figure \ref{Fig:KOH}). In either case, once \ce{KCl} clouds begin to form, abundances of both \ce{KCl} and \ce{KOH} quickly drop above the cloud deck, resulting in the \ce{KCl} cloud profile being the same in either case, except for the slight shift in the cloud deck.

For the purposes of this paper where we only consider chemical and phase equilibrium cases, we take this to be sufficient indication that there is only a minor impact from ignoring \ce{KOH}. Non-equilibrium models may be impacted, but that is beyond the scope of this paper.

\section{Convergence and Conservation}
\label{Apdx:Conv}

\begin{figure*}
	\includegraphics[]{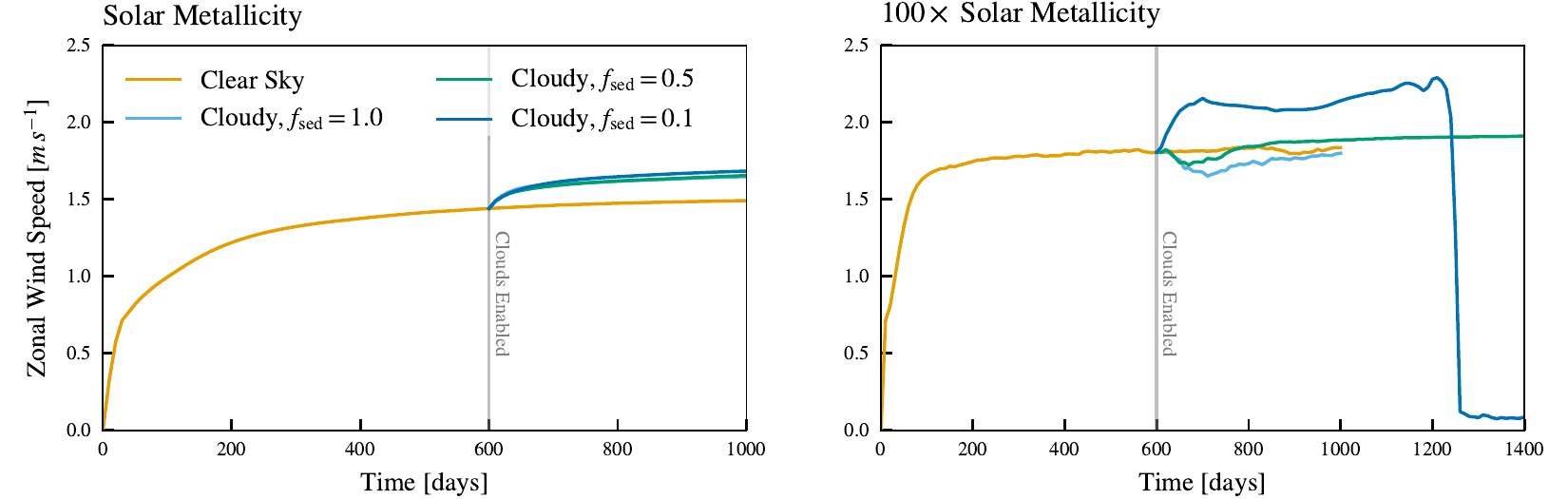}
    \caption{The evolution of the zonal velocity in each of the simulations.  For each cloudy simulation, the clouds were enabled after 600 days. The transition in the flow in the $100\times$ solar metallicity $f_\mathrm{sed}=0.1$ case can be clearly seen in the right plot.}
    \label{Fig:ZonalVelConv}
\end{figure*}

In this section, we look at the convergence of the simulations.   Examining first at the evolution of the peak zonal velocities (Figure \ref{Fig:ZonalVelConv}), the clear sky simulations quickly approach a near-steady state within the first few hundred days.  There remains a gradual continued spin-up of the atmosphere as the models lack any explicit physically-motivated dissipation mechanism.  While it is possible that the simulations may eventually reach a state in which the numerical dissipation balances the physical forcing, previous experiments (not shown) have found that the simulations become unstable before this occurs.

For the solar metallicity case, the introduction of cloud at 600 days results in a shift in the zonal velocities of $\sim 0.2 \,\mathrm{km\,s^{-1}}$ with the simulations quickly reaching a new near-steady state.  The $100\times$ solar metallicity case, however, sees fluctuations in zonal velocity with time after the introduction of clouds at 600 days, although the zonal velocities do not exhibit a long term diverging trend.   

\begin{figure*}
	\includegraphics[]{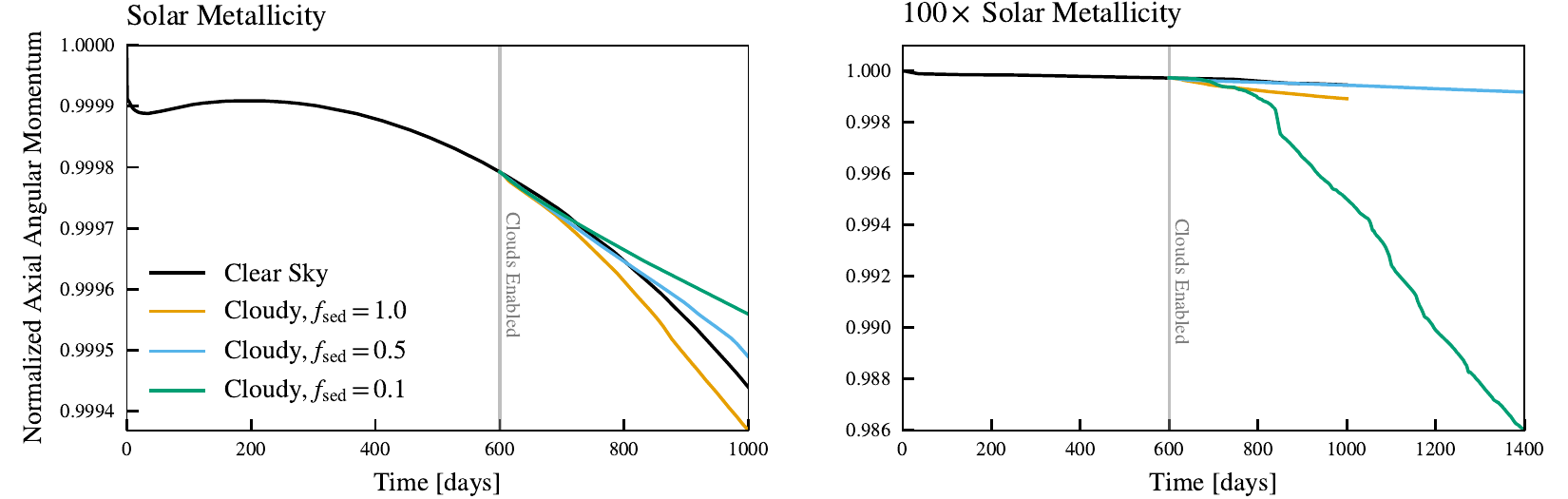}
    \caption{The evolution of the axial angular momentum within the computational domain, normalised to the initial value, for each simulation presented in the paper. }
    \label{Fig:AAM}
\end{figure*}

Although the {\sc UM} does not explicitly conserve axial angular momentum (to machine accuracy), we see it remains relatively well conserved for the solar metallicity simulations, losing $\sim 0.05\%$ over the 1000 days of simulation (see Figure \ref{Fig:AAM}, left panel).   The $100\times$ solar metallicity simulations, on the other hand, show a larger loss of angular momentum after clouds are enabled, with the $f_\mathrm{sed}=0.1$ case losing $\sim 1.4\%$ of the initial angular momentum and the majority of that loss occurring during the final 800 days (see Figure \ref{Fig:AAM}, right panel). This is possibly, in part, due to the poor conservation during the contraction of the atmosphere as well as the polar diffusion scheme dissipating angular momentum from the atmosphere.  This will be investigated in a future work. 

The transition between the flow structures in the $100\times$ solar metallicity, $f_\mathrm{sed}=0.1$ case can be seen in Figure \ref{Fig:ZonalVelConv} (right panel), with the transition in the flow strucutre being illustrated in Figure \ref{Fig:Trans2}.

\section{Additional Plots}
\label{Apdx:Plots}

\begin{figure*}
	\includegraphics[]{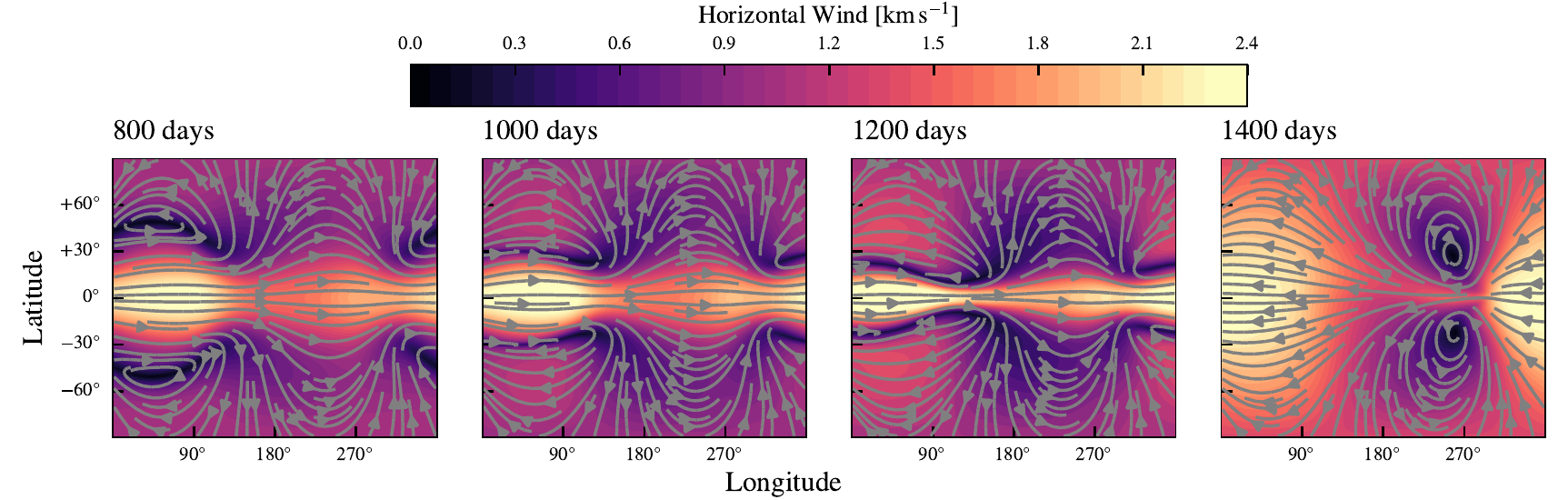}
    \caption{The evolution of the $100\times$ solar metallicity $f_\mathrm{sed}=0.1$ simulation at 0.001 bar over the length of the period in which clouds are included. }
    \label{Fig:Trans1}
\end{figure*}

\begin{figure*}
	\includegraphics[]{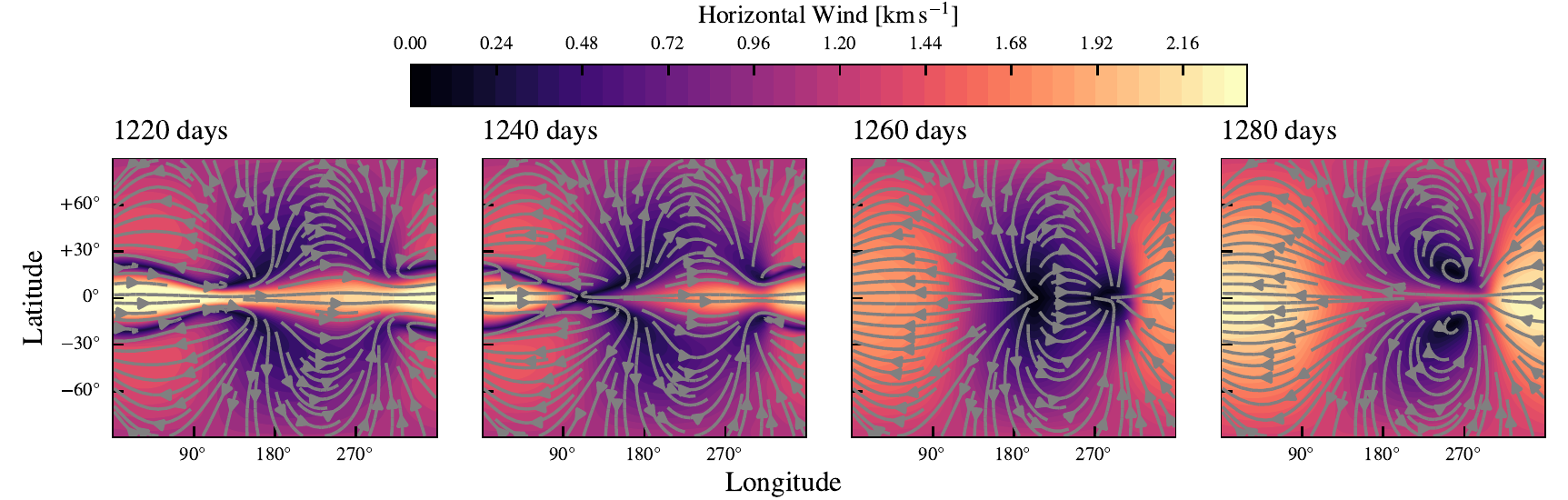}
    \caption{The same as in Figure \ref{Fig:Trans1}, except for the 60 days around the transition in the flow structure. }
    \label{Fig:Trans2}
\end{figure*}

In this appendix we include a number of additional plots illustrating the transition in flow structure seen in the $100\times$ solar metallicity $f_\mathrm{sed}=0.1$ simulation.  Figure \ref{Fig:Trans1} shows the evolution of the flow over the length of the simulation beginning at when clouds are enabled.  Figure \ref{Fig:Trans2} shows evolution over the period between 1220 days and 1280 days when the flow undergoes the largest change in structure.

\bsp	
\label{lastpage}
\end{document}